\documentclass[aps,prl,twocolumn,preprintnumbers,amsmath,amssymb,superscriptaddress]{revtex4-2}

\usepackage{times}
\usepackage{easybmat}
\usepackage{graphicx}
\usepackage{dcolumn}
\usepackage{bm}
\usepackage{color}
\definecolor{LinkColor}{rgb}{0.75,0.0,0.2}
\usepackage{verbatim}
\usepackage{hyperref}
\hypersetup{
	pdfauthor={good guys},
	pdftitle={good title},
	colorlinks=true,
	citecolor=LinkColor,
	linkcolor=LinkColor,
	urlcolor=LinkColor,
}

\begin{document}

\title{The Dual Role of Low-Weight Pauli Propagation: A Flawed Simulator but a Powerful Initializer for Variational Quantum Algorithms}

\author{Zong-Liang Li}
\affiliation{Institute of Physics, Chinese Academy of Sciences \& Beijing National Laboratory for Condensed Matter Physics, Beijing 100190, China}
\affiliation{School of Physical Sciences, University of Chinese Academy of Sciences, Beijing 100049, China}

\author{Shi-Xin Zhang}
\email{shixinzhang@iphy.ac.cn}
\affiliation{Institute of Physics, Chinese Academy of Sciences \& Beijing National Laboratory for Condensed Matter Physics, Beijing 100190, China}

\date{\today}

\begin{abstract}
Variational quantum algorithms are often hindered by rugged optimization landscapes. In this Letter, we investigate the low-weight Pauli propagation (LWPP) algorithm and find that it serves as an unreliable energy estimator for variational circuits. However, we reveal a counterintuitive insight: the Pauli-weight truncation acts as a spectral filter, effectively smoothing out high-frequency local minima while preserving the global basin of attraction in the landscape. We identify this mechanism as landscape alignment, where the approximate landscape becomes a superior navigator compared to the rugged exact landscape. Benchmarks across diverse spin models and molecular systems demonstrate that LWPP-initialized optimization yields order-of-magnitude improvements in accuracy, often finding solutions inaccessible to direct exact optimization. This work reframes LWPP from a flawed simulator into a vital pre-optimizer that serves not only as a cheap classical substitute but also as an essential tool for addressing quantum optimization challenges.

\end{abstract}

\maketitle

\textit{Introduction---}Variational quantum algorithms (VQAs) represent a leading strategy for harnessing near-term quantum devices~\cite{Preskill2018quantum, Peruzzo2014, Farhi2014, kandala2017hardware,Li2017control,Li2017bvqem, Liu2018born, Yuan2019dynamics, Khatri2019compiling, Havlicek2019nature, Beer2020dnn, cerezo2021variational,Bharti2021nisqreview, tilly2022variational,zhang2022variational, Stokes2020qng, Chen2023, liu2023probing, Miao2023nnvqa, Zhang2025postselection, Huang2023review, Cheng2024darbo,Shang2023sh, Zhang2023vqnhem, Chen2025unlearning, Zhang2025unified, Zhang2025qec}. As hybrid quantum-classical methods, VQAs employ a parameterized quantum circuit to prepare a trial state and a classical optimizer to iteratively tune the circuit parameters, seeking to minimize a given cost function. The VQA optimization landscapes are notoriously difficult, plagued by barren plateaus~\cite{McClean2018, ostaszewski2021structure,
grant2019initialization, arrasmith2021effect, wang2021noise, cerezo2021higher, marrero2021entanglement, cerezo2021cost, uvarov2021barren,
pesah2021absence, Liu2022bpvqe, Liu2022bptn, arrasmith2022equivalence, holmes2022connecting, holmes2021barren, kim2021universal, Liu2023ra, zhang2024absence, Larocca2025bpreview} and extensive local minima~\cite{Bittel2021np, Anschuetz2022traps}. This places immense importance on the choice of initial parameters, as a poor starting point can derail the entire optimization process.

Recent theoretical breakthroughs in the low-weight Pauli propagation (LWPP) algorithm~\cite{Angrisani2025, bermejo2024quantum} have shown that for typical random quantum circuits, expectation values can be estimated efficiently. These results, often based on average-case analysis over an ensemble of circuits~\cite{Nemkov2023, Aharonov2023, fontana2025classical, rudolph2023classical, Gonzlez-Garca2025quantum, Shao2024, Angrisani2025, lerch2024efficient,  angrisani2025simulating, beguvsic2025simulating, Martinez2025, rall2019simulation, bermejo2024quantum}, suggest the possibility of replacing quantum hardware with classical simulators for VQA evaluation.
However, there is a fundamental difference between these theoretical frameworks and the reality of a VQA optimization. A VQA trajectory does not explore a random ensemble of circuits. Instead, its parameters evolve via directed updates, and average-case guarantees of classical simulability do not automatically apply to such a specific, non-random instance. This raises a critical and practical question: what is the true utility of classical algorithms like LWPP in the context of VQA workflows?

In this Letter, we first demonstrate that LWPP is a poor estimator for the energy evaluation of VQAs, highlighting the limitations of applying average-case theories to single instances. However, from this failure, we uncover a counterintuitive inversion of this limitation: the numerical inaccuracy of LWPP is the very source of its optimization power. We demonstrate that the Pauli-weight truncation inherent to LWPP acts as a spectral filter, effectively smoothing out the high-frequency local minima that plague the exact quantum landscape. As a result, the approximate LWPP landscape is not merely a computationally cheaper surrogate but a topologically superior navigator. It robustly guides parameters into the global basin of attraction, avoiding the traps that frequently derail direct optimization on the exact, rugged landscape.

Leveraging this insight, we shift our perspective from using LWPP as a simulator to employing it as a classical pre-optimizer. By running an initial, purely classical optimization with the LWPP cost function, we generate superior starting parameters for the main VQA optimization loop. Benchmarking this strategy on various quantum many-body models, including various lattice topologies, hierarchical ansatzes~\cite{pesah2021absence, Cong2019, Liu2023, Bravo2022,bermejo2024quantum, Cincio2008}, molecular ground states, and noisy environments, our results show an order-of-magnitude improvement in both final accuracy and convergence speed over standard heuristics~\cite{grant2019initialization}, and this strategy often uncovers high-quality solutions inaccessible to direct quantum optimization. Our work thus reframes the role of LWPP: it is a flawed simulator but a powerful initializer and superior navigator. This practical insight offers a path to significantly accelerate VQA performance by offloading the exploratory optimization phase to classical computers, thereby reducing requirements on quantum computational resources.

\textit{The low-weight Pauli propagation algorithm---}The LWPP algorithm operates in the Heisenberg picture by tracking the backward evolution of an observable $O$ under the circuit's unitary transformation $U$. The simulation begins by decomposing $O$ into a weighted sum of Pauli strings, $O = \sum_i c_i P_i$, where each $P_i$ is a tensor product of $n$ single-qubit Pauli operators $\{I, X, Y, Z\}$. It then propagates this sum gate-by-gate to compute the expectation value with respect to the initial state, $\mathrm{Tr}[\rho_{\text{initial}} (U^\dagger O U)]$, where $U$ represents the unitary evolution of the quantum circuit composed of a sequence of gates.

The complexity arises from non-Clifford rotation gates $R_G(\theta) = e^{-i\theta G/2}$, which cause a Pauli string $P$ to branch if it does not commute with the generator $G$:
\begin{equation}
\label{eq:rotation}
    R_G(\theta)[P] = \cos(\theta)P + \sin(\theta)P', \quad \text{for } [P, G] \neq 0
\end{equation}
where $P' = i[G, P]/2$. This branching leads to an exponential growth in the number of terms.

To render the simulation tractable, LWPP applies a weight truncation approximation. The weight of a Pauli string, $P = P_1 \otimes P_2 \otimes \dots \otimes P_n$, is defined as the number of its single-qubit Pauli operators $P_i$ that are not the identity operator ($I$). At each step, any string whose weight exceeds a predefined cutoff $k$ is discarded. This approximation reduces the time complexity to polynomial scaling $\mathcal{O}(d n^{k})$~\cite{Angrisani2025}, where $n$ is the number of qubits and $d$ is the circuit depth. While this introduces approximation errors, particularly when branching coefficients ($\sin \theta$) are large, it provides a tunable trade-off between accuracy and computational cost.

\textit{A flawed estimator but a useful navigator---}Recent theoretical work has established that for typical quantum circuits, particularly those exhibiting locally scrambling properties, expectation values can be efficiently estimated classically via LWPP~\cite{Angrisani2025}. The accuracy is defined on average over an ensemble of circuits with randomly drawn parameters. However, a crucial question for practical applications remains: Do these average-case guarantees translate to the optimization of a single, specific VQA instance? To directly address this gap, we evaluate LWPP not on a random circuit ensemble, but along an explicit VQA optimization trajectory.

\begin{figure}[t!]
    \centering
    \includegraphics[width=\columnwidth]{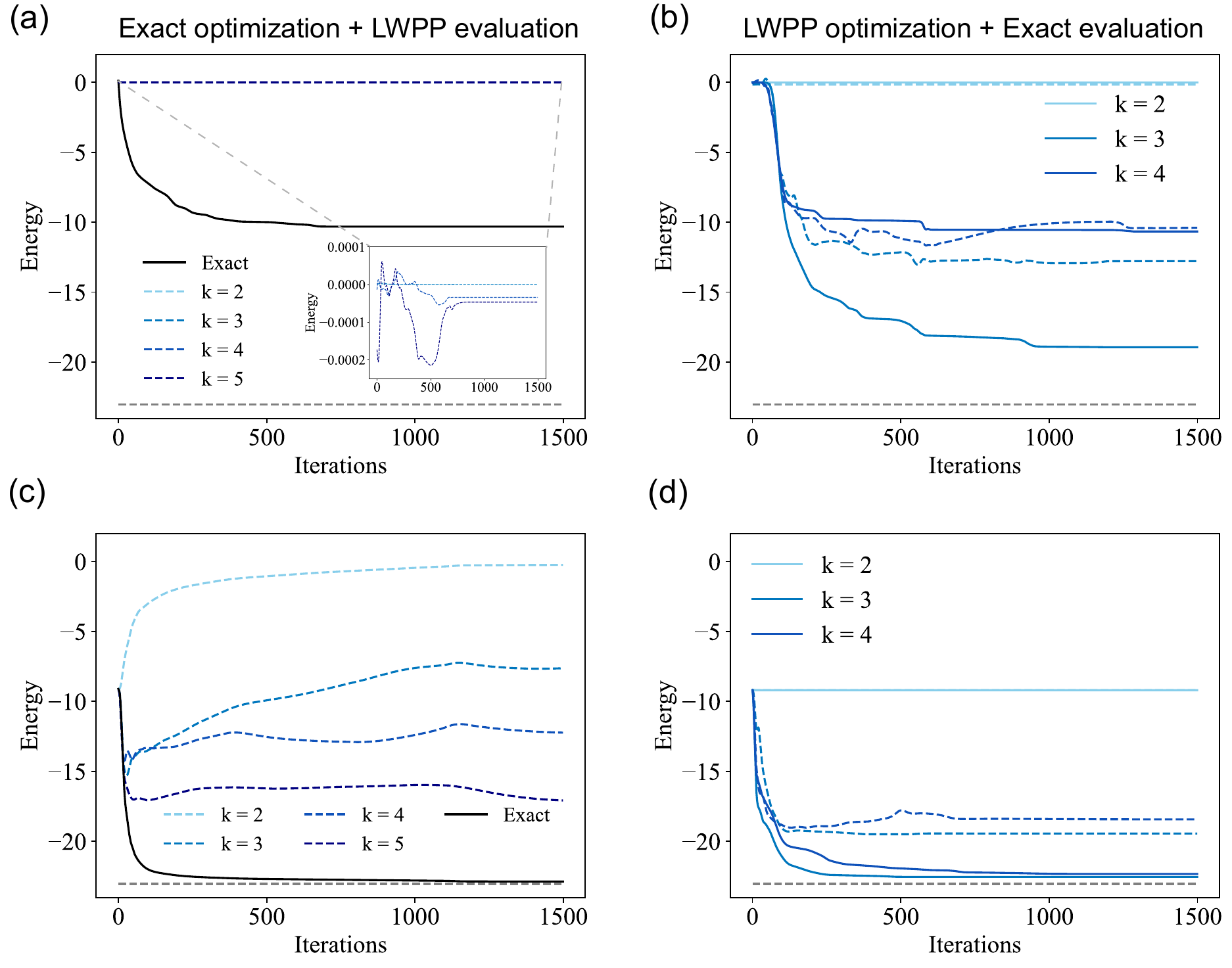}
    \caption{\textbf{Parameter-dependent accuracy of the LWPP estimation.}
    Performance analysis on a $3 \times 4$ antiferromagnetic Heisenberg XYZ model with circuit depth $d=4$ and varying truncation cutoffs $k$. The top row (a, b) starts from random initialization, while the bottom row (c, d) starts from near-identity initialization. (a, c) Exact optimization + LWPP evaluation: The optimizer is driven by the exact energy gradient $\nabla_\theta E$ (solid black line), while the LWPP energy estimate $E_{\text{LWPP}}$ (blue dashed lines) is passively recorded at each iteration to benchmark its accuracy. The inset in (a) provides a magnified view of the LWPP evaluations, which collapse to near-zero values. (b, d) LWPP optimization + Exact evaluation: The optimizer is driven by the approximate LWPP gradient $\nabla_\theta E_{\text{LWPP}}$ (solid colored lines) as the cost function. The true exact energy $E$ (dashed colored lines) is passively recorded to verify the quality of the state. Despite the numerical inaccuracy, minimizing the LWPP cost robustly guides the true energy down, suggesting its potential utility as a navigator.}
    \label{fig:accuracy_analysis}
\end{figure}

Here, we study the variational quantum eigensolver (VQE) applied to the two-dimensional Heisenberg XYZ model on a square lattice as an example,
\begin{equation}
    H = \sum_{\langle i,j \rangle} (J_x X_i X_j + J_y Y_i Y_j + J_z Z_i Z_j),
\end{equation}
where $J_x=1.0, J_y=0.8,$ and $J_z=0.5$. All simulations are implemented in {\sf TensorCircuit-NG}~\cite{zhang2023tensorcircuit}, with circuit ansatz details provided in the Supplemental Material.

We consider two complementary scenarios:  (1) Exact optimization + LWPP evaluation, where parameters are updated via the exact energy gradient while the LWPP estimate is passively recorded; and (2) LWPP optimization + Exact evaluation, where the optimizer is guided solely by the LWPP approximation while the true energy is tracked for validation. 
Both random ($\theta \in [-\pi, \pi]$) and near-identity ($\theta \in 0.01 \times [-\pi, \pi]$) initializations are examined.

The results shown in Fig.~\ref{fig:accuracy_analysis} reveal  a sharp distinction between average-case promises and single-instance performance. For random initialization, LWPP fails dramatically as a static estimator. As shown in Fig.~\ref{fig:accuracy_analysis}(a), the LWPP-evaluated energy on the true optimization path diverges from the exact value and collapses to a non-physical value near zero. This collapse occurs because the repeated truncation of terms branching from $\sin(\theta)$ (as shown in Eq.~\eqref{eq:rotation}) causes the estimate to decay exponentially with circuit depth. This confirms that LWPP cannot serve as a reliable single-instance cost estimator for VQE.

It is worth noting that this numerical failure does not preclude optimization utility. As shown in Fig. \ref{fig:accuracy_analysis}(b, d), minimizing the LWPP cost function consistently guides parameters toward regions of lower true energy. This suggests that despite its numerical inaccuracy, the LWPP landscape may capture sufficient structural features to act as an effective navigator toward high-quality basins of attraction.
This insight reframes the role of LWPP from an unreliable simulator into a potential classical pre-optimizer. Rather than seeking precise energy estimation, a task where average-case guarantees often fail for specific VQA instances, we propose leveraging the LWPP landscape to identify promising initial parameters. In the following, we benchmark this strategy on Heisenberg XYZ models to evaluate whether such pre-optimization effectively improves the optimization accuracy and efficiency.

\begin{figure}[t]
    \centering
    \includegraphics[width=\columnwidth]{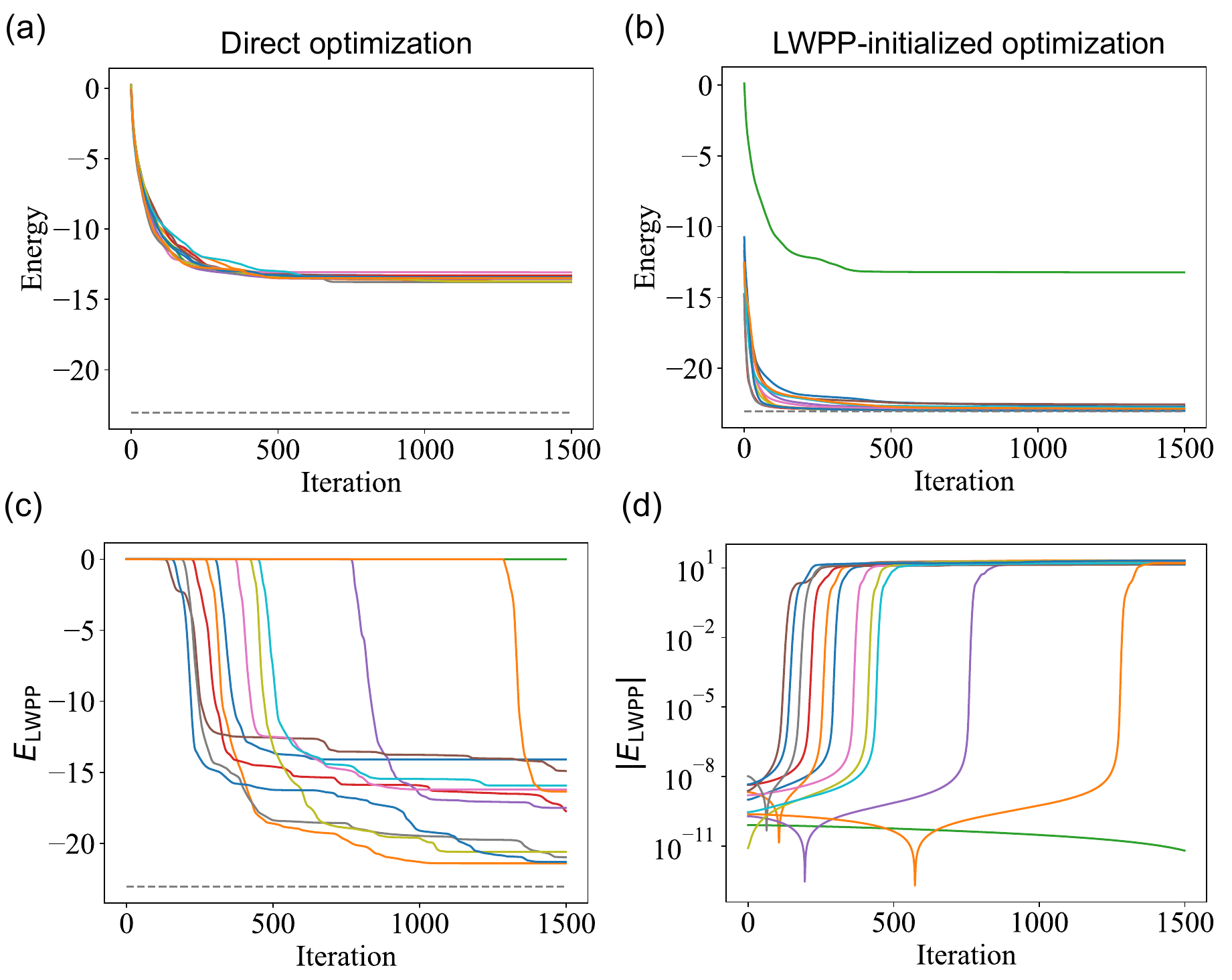}
    \caption{\textbf{An illustrative example of LWPP-initialized optimization from a random start.}
    VQA optimization of a $d=6$ VQA circuit on a $3\times4$ lattice, starting from the same random parameters. The figure contains 12 independent optimization runs, each initiated from a randomly sampled parameter set and represented by a different color.
    (a) Direct optimization consistently fails, converging to a high-energy local minimum.
    (b) LWPP-initialized optimization succeeds in finding a low-energy state.
    (c,d) The corresponding LWPP pre-optimization dynamics, with (d) showing the energy deviation from zero on a log scale.
    }
    \label{fig:rd_ex}
\end{figure}

\textit{LWPP-initialized optimization from random parameters---}To test the navigation efficacy of LWPP, we compare (1) \textbf{Direct random optimization}, which performs 1500 steps of optimization using the exact cost function, against (2) a two-stage \textbf{LWPP-initialized optimization} protocol, which consists of 1500 classical pre-optimization steps on the LWPP landscape followed by 1500 steps of exact VQA optimization. In both cases, the optimization is initiated from the same set of random parameters $\theta \in [-\pi, \pi]$.

We benchmark a $3 \times 4$ lattice (circuit depth $d=6$) in Fig.~\ref{fig:rd_ex}. While direct optimization consistently traps parameters in sub-optimal local minima [Fig.~\ref{fig:rd_ex}(a)], LWPP-initialized optimization robustly navigates to the global basin [Fig.~\ref{fig:rd_ex}(b)]. However, the internal dynamics in Fig.~\ref{fig:rd_ex}(c,d) reveal an underlying challenge: the LWPP energy signal is initially suppressed to magnitudes as low as $10^{-11}$ due to heavy truncation, creating a barren plateau-like landscape. Although the optimizer manages to escape at this intermediate depth, statistical benchmarks for $d>8$ show that this signal loss eventually becomes insurmountable as circuit depth increases (see Supplemental Material). 
This limitation suggests that a near-identity initialization could be employed, which mitigates the gradient vanishing by keeping rotation angles small. We thus evaluate whether LWPP initialization is a powerful navigator when combined with this strong heuristic.

\textit{LWPP-initialized optimization from near-identity parameters---}We now benchmark two strategies, both starting from the same near-identity parameter set ($\theta \in 0.01 \times [-\pi, \pi]$). We also compare the standard (1) \textbf{Direct near-identity optimization}, which utilizes the exact cost function immediately, against our proposed (2) \textbf{LWPP-initialized optimization}. The latter employs a two-stage protocol where near-identity parameters are first pre-optimized via the LWPP landscape to seed the subsequent exact VQA loop.

\begin{figure}[h!]
    \centering
    \includegraphics[width=\columnwidth]{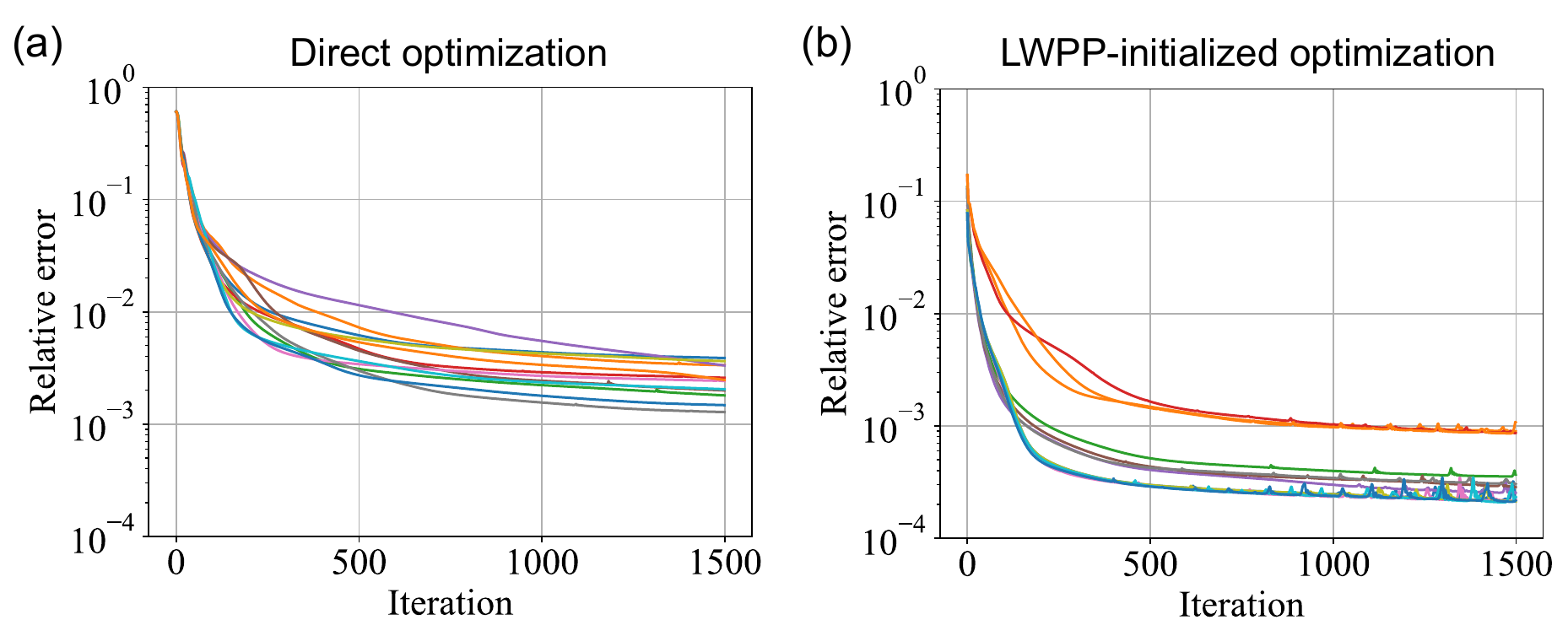}
    \caption{\textbf{A comparative example for LWPP from near-identity parameters.} Comparison of optimization trajectories of a $d=6$ circuit on a $3 \times 4$ lattice, both starting from the same near-identity parameter set. Direct optimization performs well (a), but the LWPP-initialized run (b), which includes a pre-optimization phase, achieves a significantly better final accuracy.
    }
    \label{fig:nz_single_case}
\end{figure}

We first examine a representative example on a $3 \times 4$ antiferromagnetic Heisenberg XYZ model as shown in Fig.~\ref{fig:nz_single_case}. As expected, the direct near-identity optimization performs well, achieving a final relative error of approximately $10^{-3}$ after 1500 iterations. However, the LWPP-initialized strategy proves to be even more effective. The initial LWPP pre-optimization (with $k=3$) quickly finds a parameter set with a relative error of around 10\%. Starting from this improved point, the subsequent exact optimization converges to a final relative error of nearly $10^{-4}$. This order-of-magnitude enhancement demonstrates that even when starting from a good heuristic, the LWPP pre-optimization phase is still highly superior.

\begin{figure*}[t!]
    \centering
    \includegraphics[width=\textwidth]{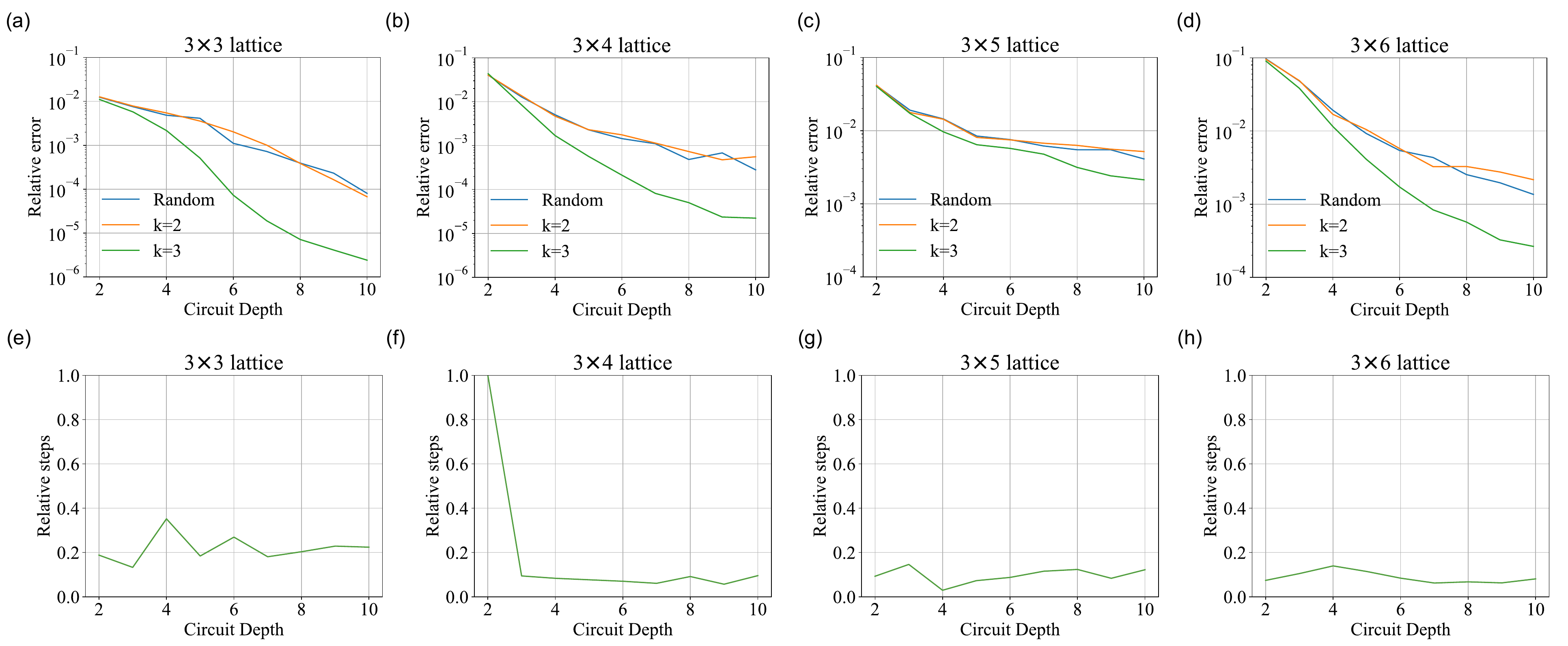}
    \caption{\textbf{Statistical comparison of initialization strategies with near-identity heuristic.} Results for antiferromagnetic models on lattices ranging from $3 \times 3$ to $3 \times 6$. (a-d) Final optimization accuracy after 1500 optimization steps: The final accuracy achieved with LWPP initialization with $k=3$ (green line) is consistently superior to direct near-identity optimization, with the advantage often increasing with circuit depth. (e-h) Optimization speedup: The relative steps required for LWPP-initialized runs ($k=3$) to reach the target accuracy, where the target is defined as the final accuracy reached by 1500 steps of direct near-identity optimization. The relative steps are calculated as the number of required iterations normalized by 1500. Lower values indicate faster convergence; LWPP-initialized optimization consistently achieves the target accuracy in a small fraction of the iterations required by the direct method, often providing a 10-fold speedup. 
    }
    \label{fig:near_zero_stats}
\end{figure*}

We then performed a statistical analysis across various lattice sizes (from $3 \times 3$ to $3 \times 6$) and circuit depths (from $d = 2$ to $d=6$), reporting the decile-based accuracy (top 10\% of runs). As shown in Fig.~\ref{fig:near_zero_stats} (a-d), LWPP-initialized optimization consistently improves the final optimization accuracy across almost all tested configurations. Notably, the margin of improvement often grows with increasing circuit depth, a regime with the emergence of barren plateaus where finding high-fidelity solutions is more challenging.

Beyond final accuracy, LWPP-initialized optimization also offers a dramatic improvement in computational speed. To quantify this, we set a target accuracy defined by the final relative error achieved after 1500 iterations of the direct near-identity optimization. We then measure the relative steps required for the LWPP-initialized runs to reach this target, where the relative steps are calculated as the number of required iterations normalized by 1500. The results, shown in Fig.~\ref{fig:near_zero_stats} (e-h), demonstrate that LWPP-initialized optimizations consistently reach the target accuracy in a small fraction of the iterations required by the direct method—often requiring only around 10\% of the steps. This represents nearly an order-of-magnitude reduction in the number of costly queries to the quantum processor. In summary, our findings consistently demonstrate that LWPP initialization provides a substantial advantage even when benchmarked against the strong near-identity initialization heuristic.

\begin{figure}[t]
    \centering
    \includegraphics[width=0.9\columnwidth]{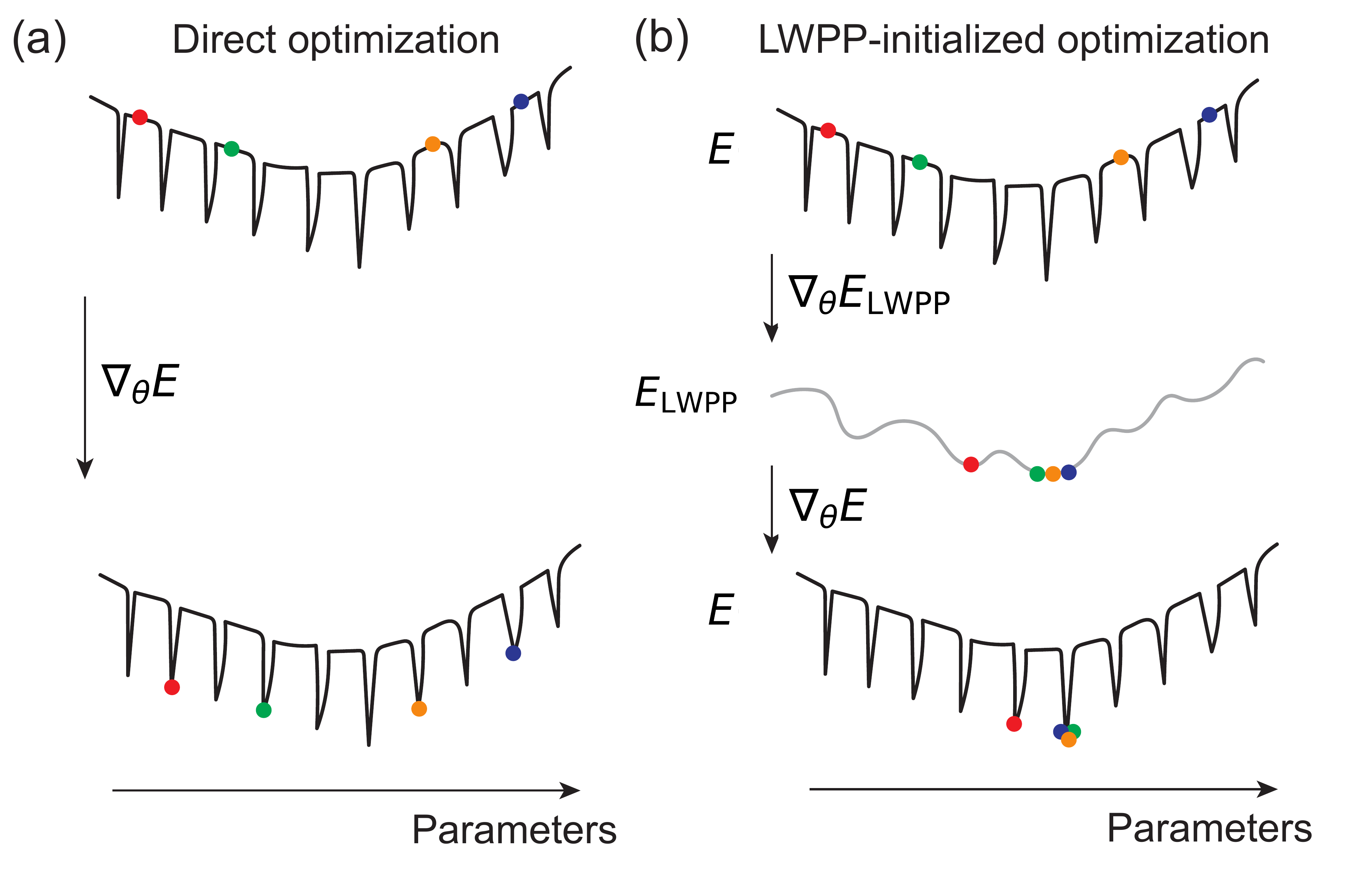} 
    \caption{\textbf{Schematic of landscape alignment.} 
    (a) Direct optimization: This optimization typically traps parameters (colored dots) in high-energy local minima.
    (b) LWPP-initialized optimization: The exact landscape (black) is rugged due to high-frequency operator spreading. The LWPP approximation (grey) acts as a smoothed surrogate by filtering out high-weight terms. Optimization on this coarse-grained landscape ($\nabla_\theta E_{\text{LWPP}}$) avoids narrow traps and robustly guides parameters into a promising basin of attraction.}
    \label{fig:landscape_alignment}
\end{figure}

\textit{Landscape alignment---}Given the remarkable and consistent improvements observed across various scenarios, a fundamental question arises regarding the mechanism behind this success. Here we propose a physical picture to explain the empirical success as landscape alignment.
Standard VQA optimization landscapes are characterized by severe ruggedness. As illustrated in Fig.~\ref{fig:landscape_alignment}(a), the exact energy landscape is densely populated with high-energy local minima. This ruggedness fundamentally arises from the scrambling nature of deep quantum circuits, where the Heisenberg-evolved observable expands into a sum of exponentially many Pauli strings. High-weight Pauli terms, resulting from extensive operator scrambling, carry high-frequency parameter dependencies that generate narrow local traps. Consequently, direct optimization ($\nabla_\theta E$) typically leaves parameters trapped in these sub-optimal valleys.

LWPP instead acts as a spectral filter by truncating terms with higher Pauli weight $k > k_{\text{cutoff}}$. LWPP implicitly discards the fine-grained ruggedness, yielding a smoothed surrogate ($E_{\text{LWPP}}$) of the true energy landscape.
As illustrated in Fig.~\ref{fig:landscape_alignment}, this filtering mechanism fundamentally alters the optimization dynamics. By removing the shallow local minima, the smoothed landscape $E_{\text{LWPP}}$ robustly guides the parameters into the broad basin of attraction of the global minimum. 

To substantiate this conceptual framework, we performed additional numerical investigations as detailed in the Supplemental Material. Through a basin probing experiment, we observed that the high-quality optima of the exact landscape are indeed located deep within the basin of the LWPP landscape. Furthermore, we demonstrated that this navigation capability is unique to the LWPP approximation. A comparison with other classical approximations, such as matrix product states (MPS), shows that classical tractability alone does not guarantee a favorable optimization landscape: unlike LWPP, MPS pre-optimization fails to guide parameters toward better basins in our benchmarks (see Supplemental Material).

\textit{Generality and robustness---}To establish the universality of our strategy, we performed extensive validations across diverse problem domains (see Supplemental Material for details). Beyond standard square lattices, we extended our benchmarks to the honeycomb lattice and other spin models as well as to molecular systems, confirming that the method's efficacy is independent of specific quantum many-body models. Notably, for hierarchical barren-plateau-free ansatzes like the multi-scale entanglement renormalization ansatz~\cite{Cincio2008}, LWPP-initialized optimization still achieved a 5-fold reduction in optimization steps. Furthermore, under realistic noise where direct gradients are often obscured, the coarse-grained LWPP landscape preserved the navigation signal, maintaining a significant advantage over direct optimization. The consistent superior results observed across such a broad spectrum of scenarios strongly suggest that these successes are universally rooted in the mechanism of landscape alignment, where the smoothed landscape consistently aligns with the true global basin regardless of model details.

Finally, we emphasize the non-triviality of our strategy with two key findings. First, the pre-optimization captures essential high-dimensional parameter correlations, rather than merely identifying a favorable independent probability distribution. As rigorously verified in the Supplemental Material, re-initializing the optimization with parameters randomly resampled from the pre-optimized distribution leads to a substantial degradation in final accuracy. 
Second, and most critically, we reveal that the approximate LWPP landscape is topologically superior to the exact quantum landscape for navigation. Rather than merely serving as a computationally cheaper classical proxy, the smoothed landscape actively filters out high-frequency local minima that entrap direct quantum optimization. This is evidenced by the fact that a hybrid protocol (e.g., 500 LWPP classical steps followed by 500 quantum steps) consistently yields lower energy solutions than a direct optimization of 1000 exact quantum steps. These results corroborate the physical picture of landscape alignment: the inaccurate classical surrogate effectively smooths the rugged optimization path, guiding the system into a global basin of attraction that is often inaccessible via the exact quantum landscape alone.

\textit{Conclusion and discussion---}In this Letter, we have established a paradoxical yet powerful role for the LWPP algorithm in VQA scenarios. We demonstrated that despite its unreliability for directly simulating the VQA cost function, LWPP serves as an effective tool for pre-optimizing VQA parameters, leading to significant improvements in both final optimization accuracy and convergence speed.

The success of this strategy can be attributed to the LWPP optimization landscape. The landscape is not numerically exact, but effectively guides parameters towards favorable basins of attraction, providing a high-quality starting point. We anticipate that this initialization method can become a standard and powerful tool in practical VQA workflows. By offloading the difficult exploratory phase of optimization to an efficient classical algorithm, LWPP-initialized optimization significantly reduces quantum-processor calls while yielding superior optimization accuracy, making the path to practical quantum advantage more attainable.

\textbf{Acknowledgments.} We acknowledge helpful discussion with Shuo Liu and Hao-Kai Zhang. SXZ acknowledges the support
from Quantum Science and Technology-National Science and Technology Major Project (No. 2024ZD0301700)
and the National Natural Science Foundation of China (No. 12574546).

%

\clearpage
\newpage
\widetext

\begin{center}
\textbf{\large Supplemental Material for ``The Dual Role of Low-Weight Pauli Propagation: A Flawed Simulator but a Powerful Initializer for Variational Quantum Algorithms''}
\end{center}

\tableofcontents

\renewcommand{\thefigure}{S\arabic{figure}}
\setcounter{figure}{0}
\renewcommand{\theequation}{S\arabic{equation}}
\setcounter{equation}{0}
\renewcommand{\thesection}{\Roman{section}}
\setcounter{section}{0}
\setcounter{secnumdepth}{4}

\section{Landscape Alignment: Picture and Validation}
\label{sec:mechanism}

\subsection{Basin Probing}
\label{subsec:basin_probing}
To empirically validate the physical picture of landscape alignment, we tracked the optimization trajectories to probe the correlation between the exact landscape $E$ and the smoothed surrogate $E_{\text{LWPP}}$. We performed a comparative study using the antiferromagnetic Heisenberg XYZ model on a $3 \times 4$ lattice with a circuit depth of $d=6$. We designed a continuous 1200-step optimization protocol to test both the guidance capability and the basin alignment.

\begin{figure*}[htbp]
    \centering
    \includegraphics[width=1.0\linewidth]{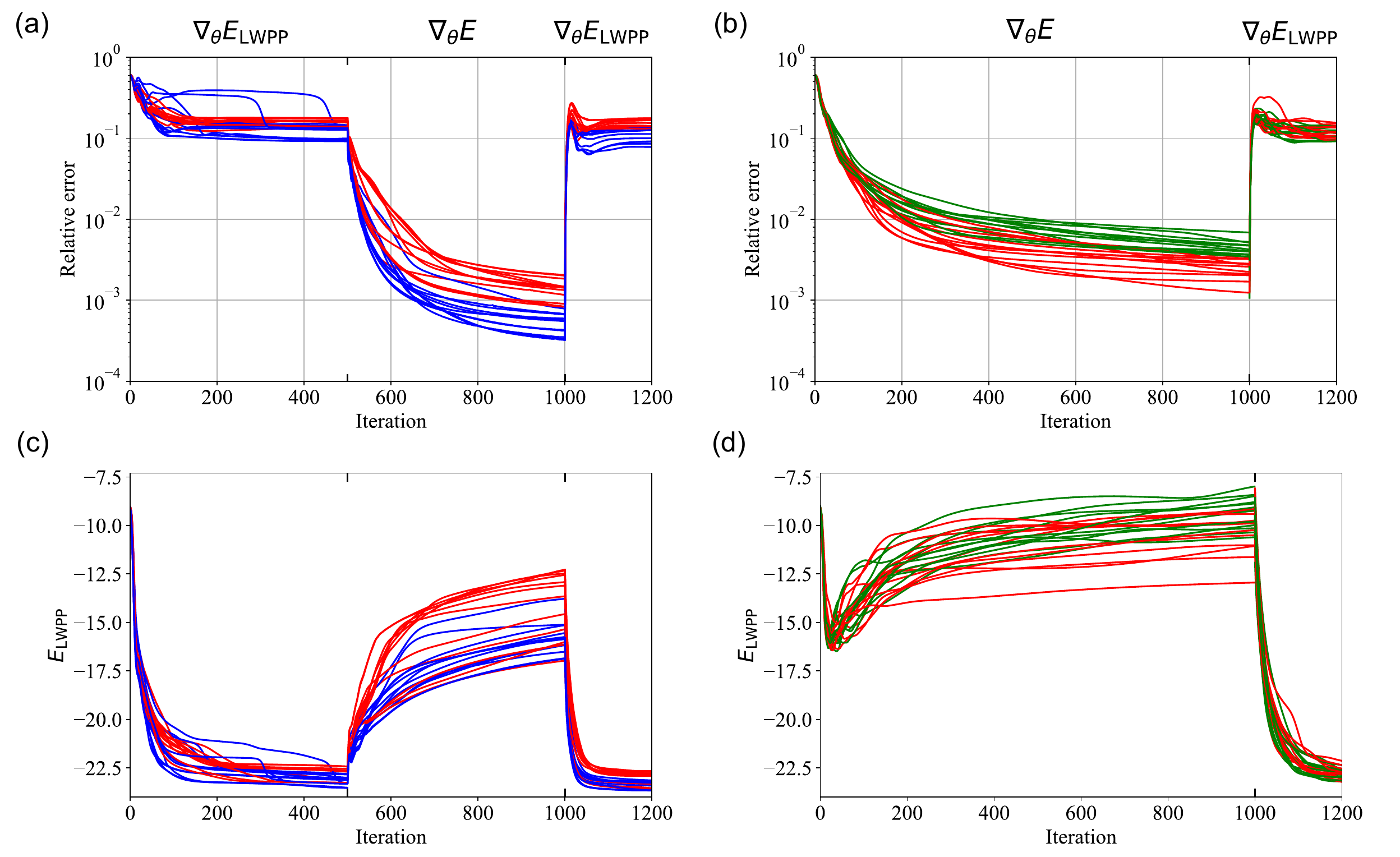} 
    \caption{\textbf{Numerical validation of landscape alignment via continuous trajectory tracking.} 
    Optimization dynamics for the antiferromagnetic Heisenberg XYZ model ($3 \times 4$ lattice, $d=6$) over 1200 iterations. The trajectories are colored by splitting independent runs into the better-performing (top 50\%) and the worse-performing (bottom 50\%) halves based on the final relative error.
    \textbf{(a, c) LWPP-initialized optimization:} 
    The optimization is divided into three phases guided by different gradients: 
    (1) \textbf{Pre-optimization} ($0\sim500$ steps) guided by $\nabla_\theta E_{\text{LWPP}}$; 
    (2) \textbf{Fine-tuning} ($500\sim1000$ steps) guided by $\nabla_\theta E$; 
    (3) \textbf{Basin probing} ($1000\sim1200$ steps) guided by $\nabla_\theta E_{\text{LWPP}}$.
    Panel (a) shows the relative error of the exact energy and Panel (c) shows the value of $E_{\text{LWPP}}$. The successful runs (Blue) achieve low relative error in phase 2 and immediately reconverge to the global minimum of $E_{\text{LWPP}}$ in phase 3.
    \textbf{(b, d) Direct optimization:}
    The optimization has two phases:
    (1) \textbf{Direct optimization} ($0\sim1000$ steps) guided by $\nabla_\theta E$; 
    (2) \textbf{Basin probing} ($1000\sim1200$ steps) guided by $\nabla_\theta E_{\text{LWPP}}$.
    As shown in panel (d), the runs that achieve lower exact energy (Red) also tend to reside in regions with lower $E_{\text{LWPP}}$ values compared to the worse-performing runs (Green). However, even these better-performing direct runs remain trapped in local minima with $E_{\text{LWPP}}$ values significantly higher than those reached by the LWPP-initialized strategy (compare Phase 3 in panel c), confirming that direct optimization fails to navigate into the global optimal basin.}
    \label{fig:landscape_validation}
\end{figure*}

The results are presented in Fig.~\ref{fig:landscape_validation}. The left column (a, c) illustrates the LWPP-initialized optimization strategy, while the right column (b, d) represents the direct optimization strategy.

For the LWPP initialization strategy, the process is divided into three distinct phases, marked by the driving gradients:
\begin{enumerate}
    \item \textbf{Pre-optimization ($0\sim500$ steps, $\nabla_\theta E_{\text{LWPP}}$):} Parameters are optimized on the smoothed landscape. As seen in Fig.~\ref{fig:landscape_validation}(c), the LWPP energy decreases rapidly to $\approx -22.5$, implying the landscape for LWPP is much smoother.
    \item \textbf{Fine-tuning ($500\sim 1000$ steps, $\nabla_\theta E$):} The optimizer switches to the exact energy gradient. Remarkably, as shown in Fig.~\ref{fig:landscape_validation}(a), the relative error for the blue group plunges to $10^{-3}$, indicating the parameters were already in a high-quality basin. Interestingly, in Fig.~\ref{fig:landscape_validation}(c), the $E_{\text{LWPP}}$ value increases slightly during this phase, reflecting the slight shift from the approximate minimum to the true exact minimum.
    \item \textbf{Basin probing ($1000\sim1200$ steps, $\nabla_\theta E_{\text{LWPP}}$):} We switch guidance back to $\nabla_\theta E_{\text{LWPP}}$. The blue trajectories in Fig.~\ref{fig:landscape_validation}(c) immediately drop back to the deep minimum. This ``round-trip'' success confirms that the exact global solution is topologically connected to and located deep within the optimal basin of the LWPP landscape.
\end{enumerate}

In contrast, the Direct optimization strategy reveals a different topology:
\begin{enumerate}
    \item \textbf{Direct optimization ($0\sim1000$ steps, $\nabla_\theta E$):} Parameters are updated using exact gradients. As shown in Fig.~\ref{fig:landscape_validation}(b), even the best runs (Red lines) plateau at a higher relative error compared to the LWPP strategy.
    \item \textbf{Basin probing ($1000\sim1200$ steps, $\nabla_\theta E_{\text{LWPP}}$):} When we switch guidance to $\nabla_\theta E_{\text{LWPP}}$, a striking contrast appears. In Fig.~\ref{fig:landscape_validation}(d), the trajectories start this phase at a high energy level ($\approx -10$ to $-15$) and converge slowly. This indicates that the local minima found by direct optimization are located in peripheral, high-energy regions of the LWPP landscape, far from the global basin identified in Fig.~\ref{fig:landscape_validation}(c).
\end{enumerate}

These results strongly support our hypothesis: minimizing $E_{\text{LWPP}}$ acts as a robust navigator, effectively filtering out high-frequency noise and guiding the system into the correct basin of attraction, a feat that direct optimization on the rugged exact landscape fails to achieve.

\subsection{Comparison with MPS Initialization}
\label{subsec:mps_comparison}
To further demonstrate that the success of LWPP initialization stems from its unique spectral filtering property rather than merely being a generic approximate method, we compared it against the widely used matrix product state (MPS) method. MPS approximates the quantum state by truncating the bond dimension ($\chi$), a strategy fundamentally different from LWPP's Pauli weight truncation.

\begin{figure}[htbp]
    \centering
    \includegraphics[width=1.0\linewidth]{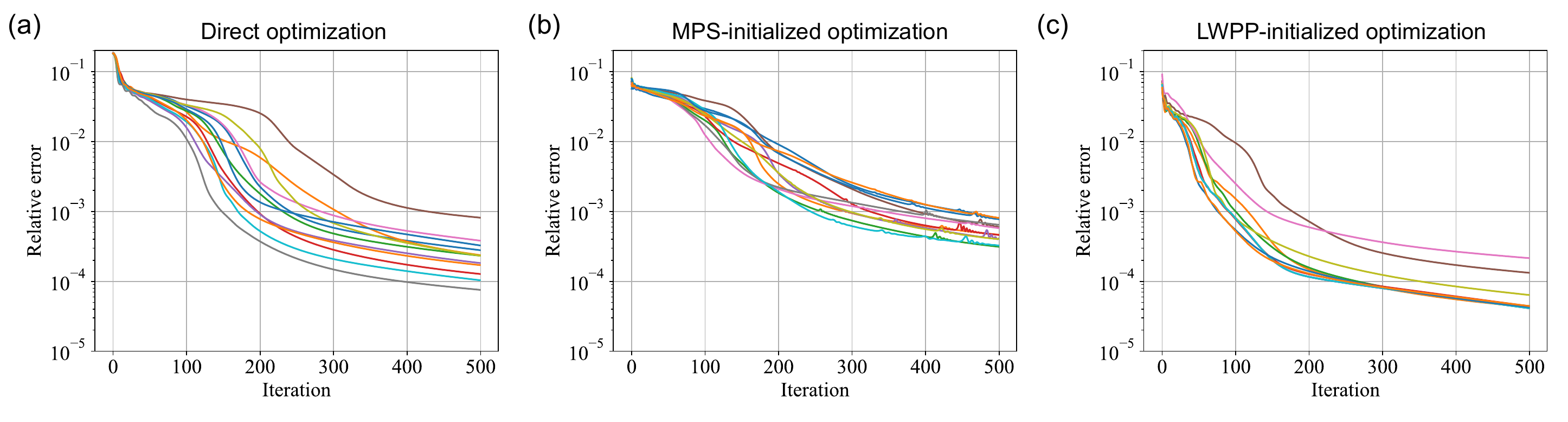} 
    \caption{\textbf{Uniqueness of LWPP initialization compared to MPS.} Optimization trajectories for a 1D Heisenberg XYZ model ($N=12, d=4$) comparing different initialization strategies.
    (a) \textbf{Direct near-identity optimization:} Standard optimization starting from near-identity parameters.
    (b) \textbf{MPS-initialized optimization:} Optimization seeded by parameters pre-optimized using an MPS approximation with bond dimension $\chi=30$. Despite the high bond dimension, the subsequent exact optimization performs poorly, indicating misalignment with the true optimization landscape.
    (c) \textbf{LWPP-initialized optimization:} Optimization seeded by LWPP pre-optimization. Crucially, while the starting energy may be comparable to other methods, the LWPP-seeded runs demonstrate a distinct capability for rapid continued convergence, confirming that LWPP uniquely locates the trainable basin of attraction.}
    \label{fig:mps_comparison}
\end{figure}

We performed a benchmark on a 1D Heisenberg XYZ model. For the MPS pre-optimization, we utilized a relatively high precision setting with a maximum bond dimension of $\chi=30$ (retaining the top 30 singular values during singular value decomposition for each application of two-qubit gates). The comparative results are shown in Fig.~\ref{fig:mps_comparison}.

\begin{itemize}
    \item \textbf{Direct optimization (Fig.~\ref{fig:mps_comparison}a):} Starting from near-identity parameters, the direct optimization achieves moderate convergence.
    \item \textbf{MPS-initialized optimization (Fig.~\ref{fig:mps_comparison}b):} Pre-optimization using the MPS cost function yields parameters that, while potentially having a lower initial energy than random guesses, fail to facilitate further optimization. The subsequent exact optimization stagnates or converges much slower than even the direct optimization. This suggests that the MPS approximation landscape, despite being a good state approximator, may possess gradients that guide parameters into spurious local minima.
    \item \textbf{LWPP-initialized optimization (Fig.~\ref{fig:mps_comparison}c):} In stark contrast, although the starting energy of the exact optimization phase might appear comparable to or slightly worse than MPS, the \textbf{subsequent convergence behavior is fundamentally different}. LWPP-initialized parameters exhibit a rapid descent, demonstrating that they are situated within a high-quality basin of attraction with favorable gradients, rather than a local trap.
\end{itemize}

This comparison highlights the uniqueness of LWPP: its truncation strategy aligns well with the global topology of the optimization landscape, whereas other approximations like MPS do not necessarily provide a favorable navigation gradient for initialization.

\subsection{Gradient Scaling Analysis}
\label{sec:gradients}
\begin{figure}[h]
    \centering
    \includegraphics[width=0.7\linewidth]{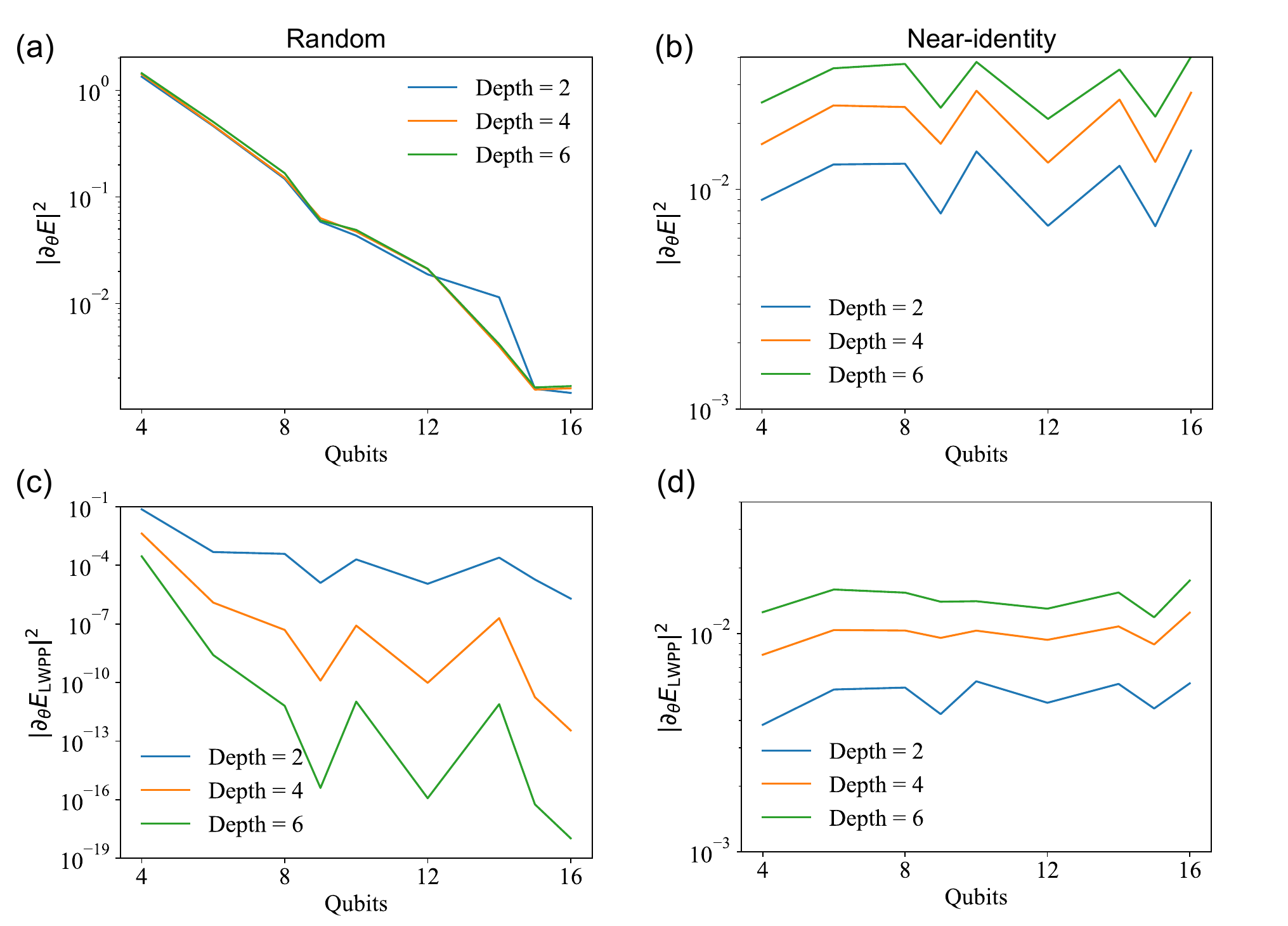} 
    \caption{\textbf{Scaling of gradient magnitudes.} The mean squared partial derivative of the energy is plotted against the number of qubits for different circuit depths ($d=2, 4, 6$).
    (a, c) Under \textbf{Random initialization}, both Exact (a) and LWPP (c) landscapes suffer from exponential gradient decay (Barren Plateaus).
    (b, d) Under \textbf{Near-identity initialization}, both Exact (b) and LWPP (d) landscapes maintain non-vanishing, trainable gradients.
    Note that the LWPP gradients (d) are not larger than the exact gradients (b); rather, the LWPP landscape is smoother. This confirms that the method's success stems from landscape smoothing (avoiding local traps) rather than gradient amplification.}
    \label{fig:gradients}
\end{figure}

To investigate whether the success of LWPP-initialized optimization stems from gradient amplification or landscape topology, we analyzed the scaling of the mean squared partial derivative, $\mathbb{E}[|\partial_\theta \mathcal{L}|^2]$, with respect to the variational parameters. We tracked this metric against the number of qubits for different circuit depths under both random and near-identity initializations.

The results, summarized in Fig.~\ref{fig:gradients}, reveal distinct behaviors for the two initialization strategies. Under random initialization, both the exact landscape (Fig.~\ref{fig:gradients}a) and the LWPP landscape (Fig.~\ref{fig:gradients}c) exhibit exponential decay of gradients, characteristic of the barren plateau problem. For LWPP, this decay arises from the compounding effect of coefficient truncation at large rotation angles, confirming that LWPP does not inherently solve the barren plateau problem in deep circuits when starting from random parameters. In contrast, the near-identity heuristic effectively resolves the vanishing gradient issue for both landscapes. As shown in Fig.~\ref{fig:gradients}(b) and (d), the gradients for both the exact energy and the LWPP energy remain significant and do not vanish exponentially in this regime.

Crucially, a comparison between Fig.~\ref{fig:gradients}(b) and (d) clarifies that the success of LWPP initialization is not due to providing larger gradients. In fact, the gradient magnitudes in the LWPP landscape are comparable to, or even slightly smaller than, those in the exact landscape. This observation supports our \textit{landscape alignment} hypothesis: LWPP acts as a spectral filter that truncates high-weight Pauli terms, thereby smoothing the high-frequency oscillations (ruggedness) of the exact landscape. While this smoothing does not amplify the gradients, it renders them topologically superior, filtering out local minima and providing a consistent direction towards the global basin of attraction. Furthermore, within the near-identity regime, we observe that the gradient magnitude actually increases with circuit depth, providing a structural clue for the enhanced optimization performance.

\section{Universality and Robustness of LWPP-initialized optimization}
\label{sec:benchmarks}

\subsection{Statistical Analysis of Random Initialization}
\begin{figure}[h]
    \centering
    \includegraphics[width=0.7\linewidth]{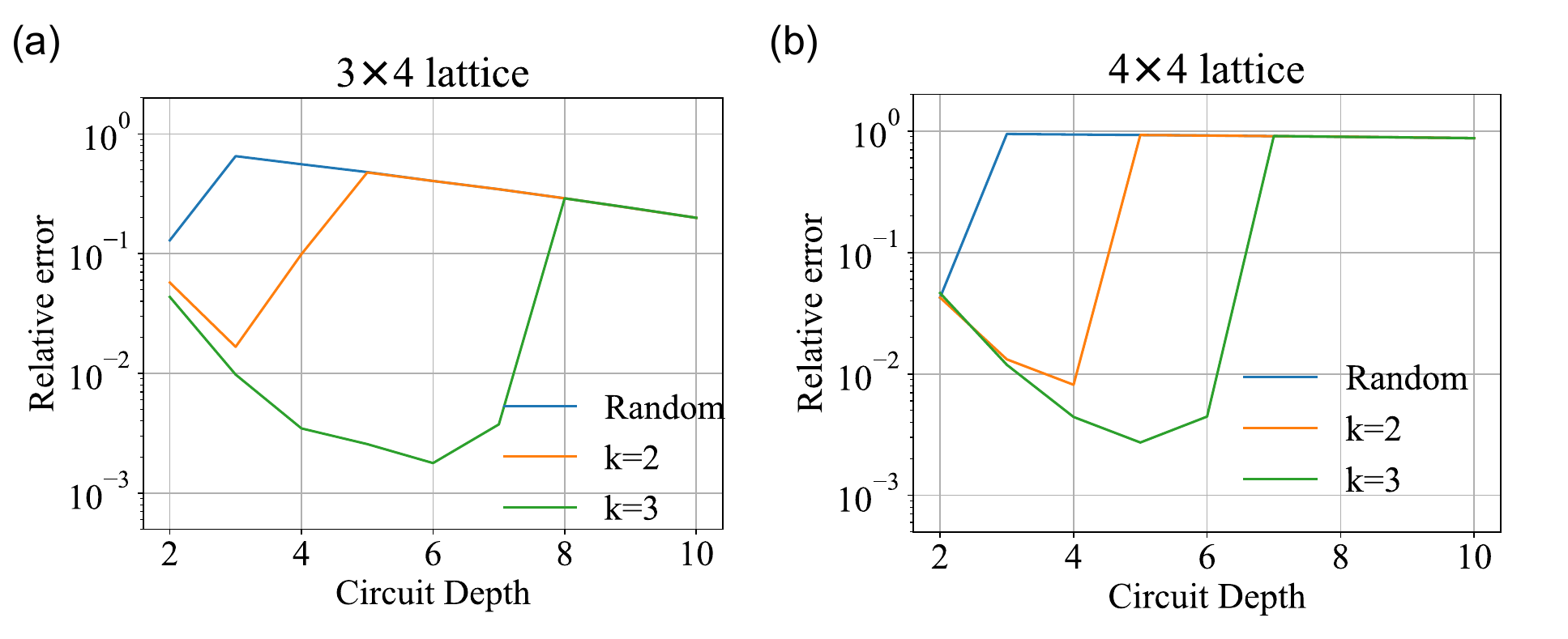}
    \caption{\textbf{Statistical comparison of initialization strategies from a random start.} Final optimization accuracy (top 10\% of runs) versus circuit depth for $3 \times 4$ (a) and $4 \times 4$ lattices (b). Direct random initialization (blue) consistently performs poorly. LWPP-initialized optimization (orange and green) achieves high accuracy for shallow and intermediate depths. However, for deeper circuits, the compounding effect of truncation on random parameters leads to a vanishing gradient landscape, reducing the effectiveness of the pre-optimization. This limitation motivates the use of near-identity initialization for deep circuits, as discussed in the main text.}
    \label{fig:rd_sta}
\end{figure}

In the main text, we demonstrated the efficacy of LWPP initialization using a representative single case for random parameters. Here, we provide a comprehensive statistical analysis of this strategy across different lattice sizes and circuit depths. We compared the \textbf{Direct random optimization} against the \textbf{LWPP-initialized optimization} (starting from random parameters $\theta \in [-\pi, \pi]$), with both methods performed over 1500 optimization steps. To provide a robust measure of performance, we report the accuracy of the top 10\% of optimization runs derived from 30 independent trials for each configuration.
The results, summarized in Fig.~\ref{fig:rd_sta}, indicate distinct behaviors depending on circuit depth. For shallow to intermediate depths, LWPP-initialized optimization (orange and green lines) significantly outperforms direct random optimization (blue line). Specifically, for the $3 \times 4$ lattice, LWPP enables the optimizer to find high-quality solutions that direct optimization completely misses. However, as the circuit depth increases (e.g., $d > 8$), the performance of LWPP initialization from random parameters begins to degrade, approaching that of the direct method.

It is worth noting that this performance drop at large depths is specific to random initialization. With large rotation angles, the number of Pauli strings generated by branching grows rapidly. Consequently, the truncation mechanism ($k_{\text{cutoff}}$) discards a significant portion of the operator weight, causing the coefficients of the surviving terms to decrease exponentially. This leads to a flattening of the LWPP landscape (analogous to a barren plateau), where the gradients become too small to guide the optimizer effectively. This observation aligns with our analysis in Section~\ref{sec:gradients}, which demonstrates that the gradient magnitudes decay exponentially with circuit depth under random initialization.

Notably, this limitation is overcome by the near-identity initialization strategy used in the main text. As shown in Section~\ref{sec:gradients}, starting near the identity preserves the gradient magnitudes even for deep circuits, ensuring that LWPP remains a powerful navigator where random initialization fails.

\subsection{Statistical Analysis of the Ferromagnetic Heisenberg XYZ Model}

\begin{figure*}[t!]
    \centering
    \includegraphics[width=\textwidth]{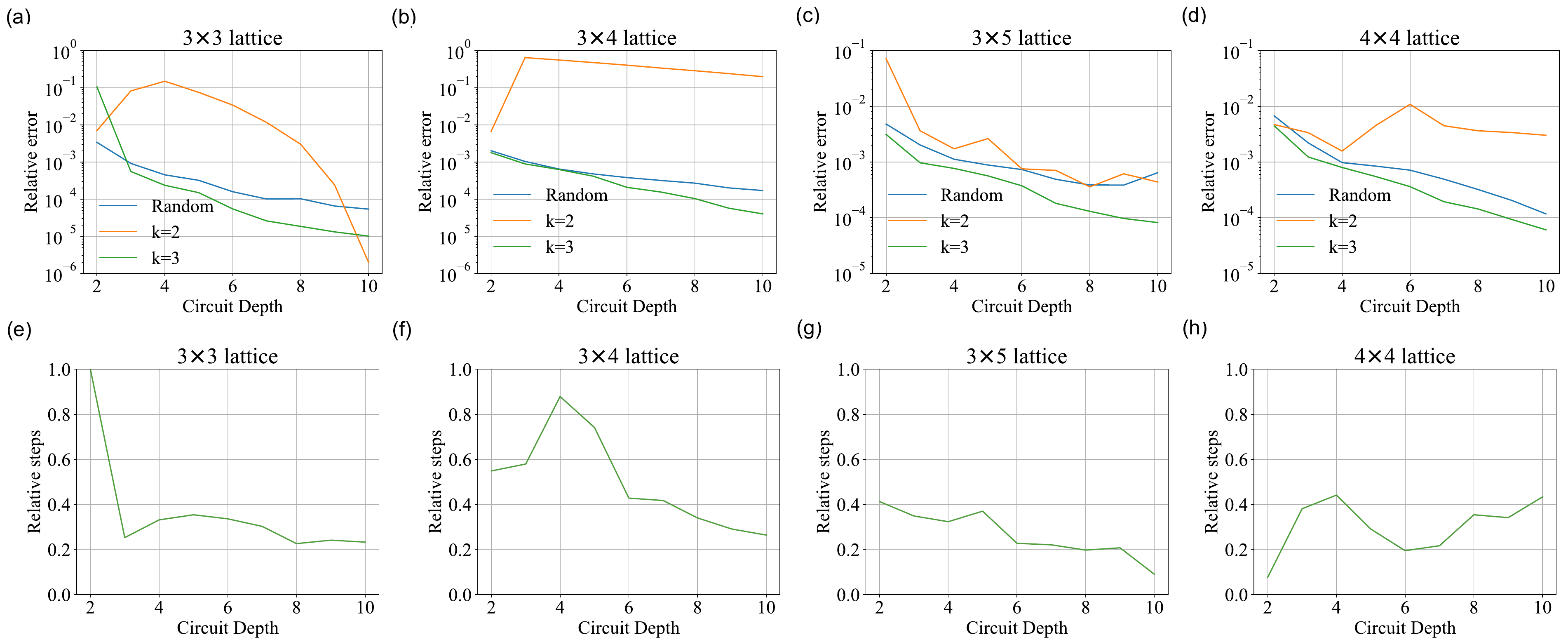}
    \caption{\textbf{Comparison of initialization strategies with near-identity heuristic on the ferromagnetic Heisenberg XYZ model.} Results are shown for various lattice sizes and number of qubits, comparing the near-identity heuristic with LWPP initialization.
    \textbf{(a-d) Final optimization accuracy:} The final accuracy achieved with LWPP-initialized optimization with $k=3$ (green line) is consistently superior to direct optimization, with the performance gap often widening with circuit depth. 
    \textbf{(e-h) Optimization speedup:} The relative steps (steps normalized by 1500) required for the LWPP-initialized runs to reach the final accuracy achieved by 1500 steps of direct near-identity optimization. Lower values signify faster convergence, with LWPP typically reaching the target in a small fraction of the iterations.
    }
    \label{fig:ferro_stats}
\end{figure*}

To test the generality and robustness of the LWPP initialization strategy, we performed an identical analysis on the ferromagnetic Heisenberg XYZ model. This model is defined by the same Hamiltonian structure as in the main text, but with ferromagnetic couplings: $J_x = -1.0$, $J_y = -0.8$, and $J_z = -0.5$. These negative couplings favor parallel spin alignment, making the model's ground state physically distinct from its antiferromagnetic counterpart. We repeated the head-to-head comparison between direct near-identity optimization and the LWPP-initialized approach across the same range of lattice sizes and circuit depths, with both strategies started from the same near-identity parameter distribution.

The results, summarized in Fig.~\ref{fig:ferro_stats}, confirm that the advantages of our method are not specific to the antiferromagnetic case. Despite the different physics, the LWPP pre-optimization provides a substantial and systematic improvement. Panels (a-d) show that LWPP-initialized optimization consistently achieves a higher final accuracy. Furthermore, panels (e-h) demonstrate a dramatic optimization speedup, with the LWPP-initialized runs reaching the target accuracy in a small fraction of the iterations on quantum hardware required by the direct method. This confirms that LWPP initialization is a powerful and versatile tool for accelerating variational quantum algorithms (VQA) across different models.

\subsection{Generalization to Hierarchical Architectures: Performance on MERA}
To address the question of whether the utility of LWPP initialization is limited to hardware-efficient ansatzes or specific square lattices, we extended our benchmarks to a fundamentally different class of variational circuits: hierarchical architectures inspired by the multi-scale entanglement renormalization ansatz (MERA)~\cite{Cincio2008}. These architectures, similar to quantum convolutional neural networks~\cite{pesah2021absence, Cong2019, Liu2023}, typically feature logarithmic depth and are known to be more robust against barren plateaus compared to standard hardware-efficient ansatzes.

We employed a 1D MERA ansatz on $N=16$ qubits. The circuit consists of $\log_2 N$ layers. In each layer, entangling gates (parameterized rotation gates generated by $XX$ and $ZZ$ interactions) are applied between qubits separated by a stride that doubles with each layer, followed by single-qubit $R_x$ and $R_z$ rotations. This structure efficiently captures long-range correlations with a shallow effective circuit depth. The target Hamiltonian is the 1D transverse field Ising model with periodic boundary conditions:
\begin{equation}
\hat{H} = J \sum_{i} Z_i Z_{i+1} - B_x \sum_{i} X_i,
\end{equation}
where we set the parameters to the critical point $J=1.0$ and $B_x=1.0$.

\begin{figure*}[htbp]
    \centering
    \includegraphics[width=0.7\linewidth]{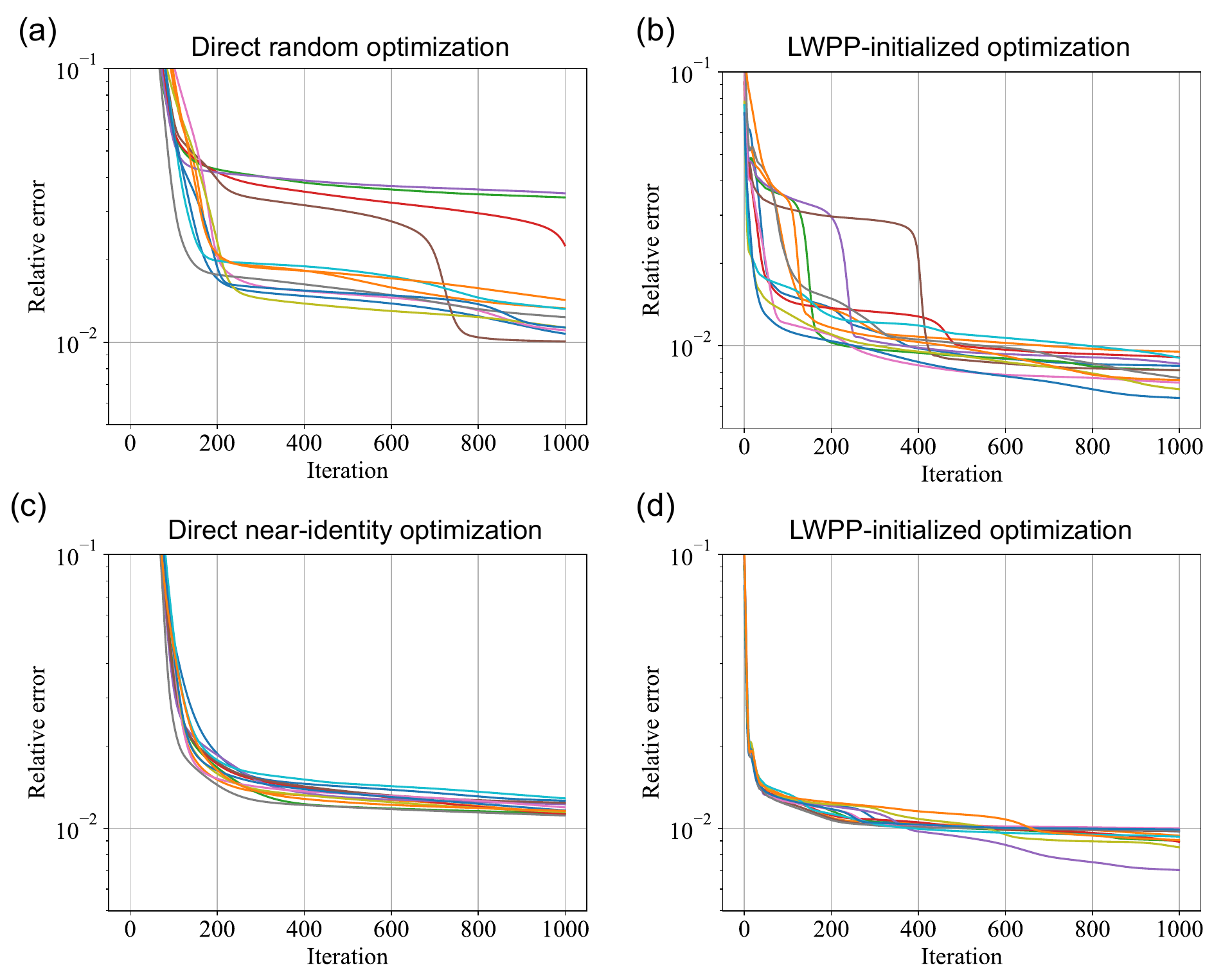}
    \caption{\textbf{Performance of LWPP-initialized optimization on a MERA-like hierarchical ansatz.} Simulations were performed on a 16-qubit 1D transverse field Ising model ($J=1, B_x=1$). The ansatz features a logarithmic depth structure with layers of increasing interaction range.
    (a, b) \textbf{Random start:} Comparison between (a) Direct optimization and (b) LWPP-initialized optimization starting from random parameters, which consistently leads to faster convergence and lower final errors.
    (c, d) \textbf{Near-identity start:} Comparison between (c) direct optimization and (d) LWPP-initialized optimization starting from near-identity parameters. Even starting from a near-identity heuristic, LWPP (d) provides a significant acceleration, achieving target accuracy in $\sim 200$ iterations compared to $>1000$ for the direct method, yielding a 5-fold reduction in optimization steps.}
    \label{fig:mera_benchmark}
\end{figure*}

The comparative results are presented in Fig.~\ref{fig:mera_benchmark}. We tested two initialization baselines: random initialization (top row) and near-identity initialization (bottom row). As shown in the figure, LWPP-initialized optimization demonstrates a clear advantage in both convergence accuracy and optimization speed across both scenarios. For instance, the LWPP-initialized runs typically reach high accuracy (relative error $< 10^{-2}$) within approximately 200 iterations, representing a roughly 5-fold reduction in optimization steps compared to the direct method, which often requires over 1000 iterations to achieve similar precision. This performance behavior matches our findings for the Heisenberg XYZ model discussed in the main text, confirming that the efficacy of LWPP initialization extends effectively to hierarchical trainable architectures.

These results demonstrate that the benefits of LWPP initialization—specifically the navigation toward high-quality basins—are not confined to barren-plateau-prone circuits. Even for inherently trainable hierarchical architectures like MERA, LWPP acts as a powerful accelerator, significantly reducing the classical-quantum feedback loops required for convergence, as it can eliminate local minima via the approximate landscape.

\subsection{Generalization to Different Lattice Topologies: Honeycomb Lattice}
To address the concern regarding whether the efficacy of LWPP initialization is tied to specific qubit connectivities, we extended our analysis to the honeycomb lattice. Unlike the square lattice (coordination number $z=4$) used in the main text, the honeycomb lattice features a lower coordination number ($z=3$) and distinct topological features. We simulated a $2 \times 3$ lattice comprising $N=12$ qubits under the ferromagnetic Heisenberg XYZ model, employing a hardware-efficient ansatz compatible with the honeycomb topology with circuit depth $d=6$.

\begin{figure}[htbp]
    \centering
    \includegraphics[width=0.7\linewidth]{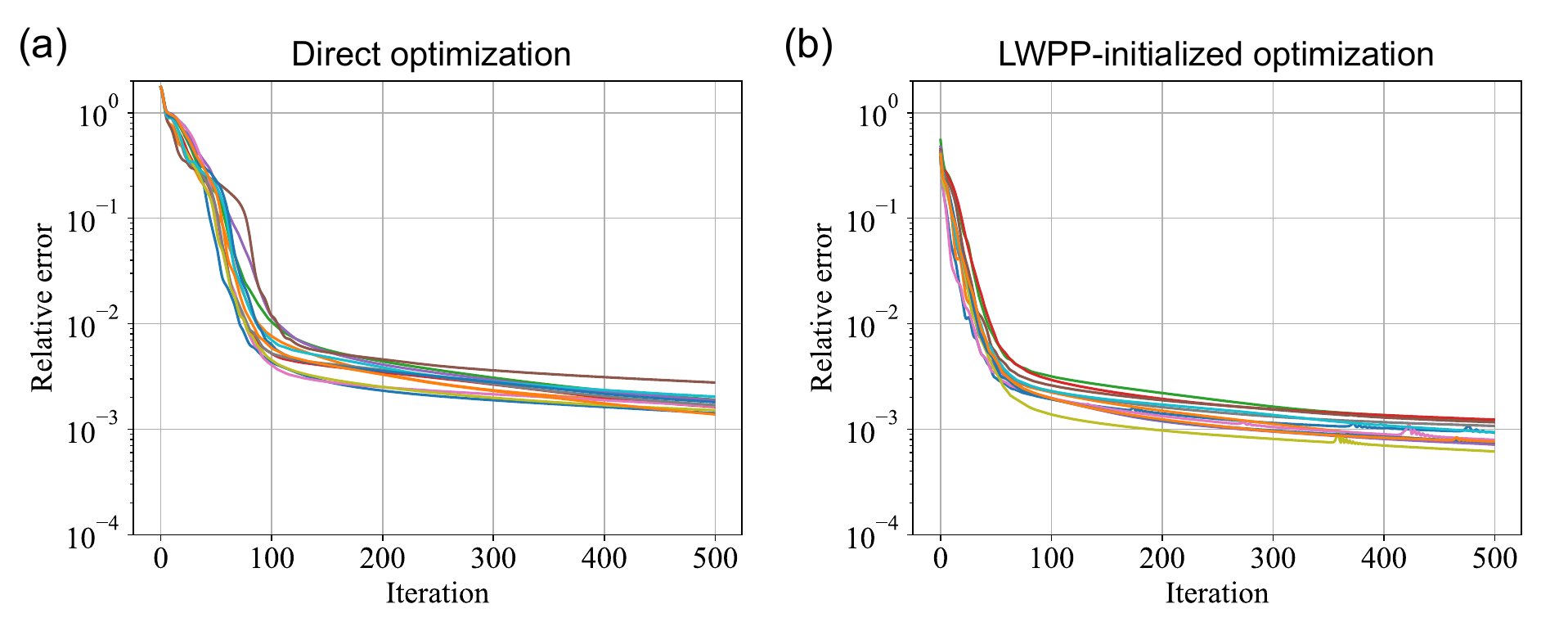} 
    \caption{\textbf{Performance on a honeycomb Lattice.} Optimization trajectories for the Ferromagnetic Heisenberg XYZ model on a $2 \times 3$ honeycomb lattice ($N=12$ qubits) with circuit depth $d=6$.
    (a) \textbf{Direct optimization:} Standard optimization starting from near-identity parameters.
    (b) \textbf{LWPP-initialized optimization:} Optimization seeded by LWPP pre-optimization. The LWPP strategy maintains its advantage on this different topology, showing faster convergence and consistently lower final energy errors compared to the direct method.}
    \label{fig:honeycomb_benchmark}
\end{figure}

The comparative results are shown in Fig.~\ref{fig:honeycomb_benchmark}. While the direct optimization starting from near-identity parameters [Fig.~\ref{fig:honeycomb_benchmark}(a)] achieves decent convergence, it exhibits a noticeable spread in final accuracies, with several trajectories getting stuck at higher energy levels. In stark contrast, the LWPP-initialized runs [Fig.~\ref{fig:honeycomb_benchmark}(b)] demonstrate superior consistency and convergence speed. Almost all trajectories rapidly converge to high-precision solutions (relative error $< 10^{-3}$), maintaining the performance advantage observed in square lattices. This confirms that the landscape smoothing mechanism of LWPP is universal in lattice geometry and connectivity.

\subsection{Generalization to Quantum Chemistry: The $\text{H}_2\text{O}$ Ground State}
To verify that the LWPP initialization method is broadly applicable, we tested it on a different type of variational quantum eigensolver (VQE) problem: finding the ground state of a molecule. The specific task was to find the ground state of a water molecule (H\(_2\)O).

\begin{figure*}[h!]
    \centering
    \includegraphics[width=0.7\textwidth]{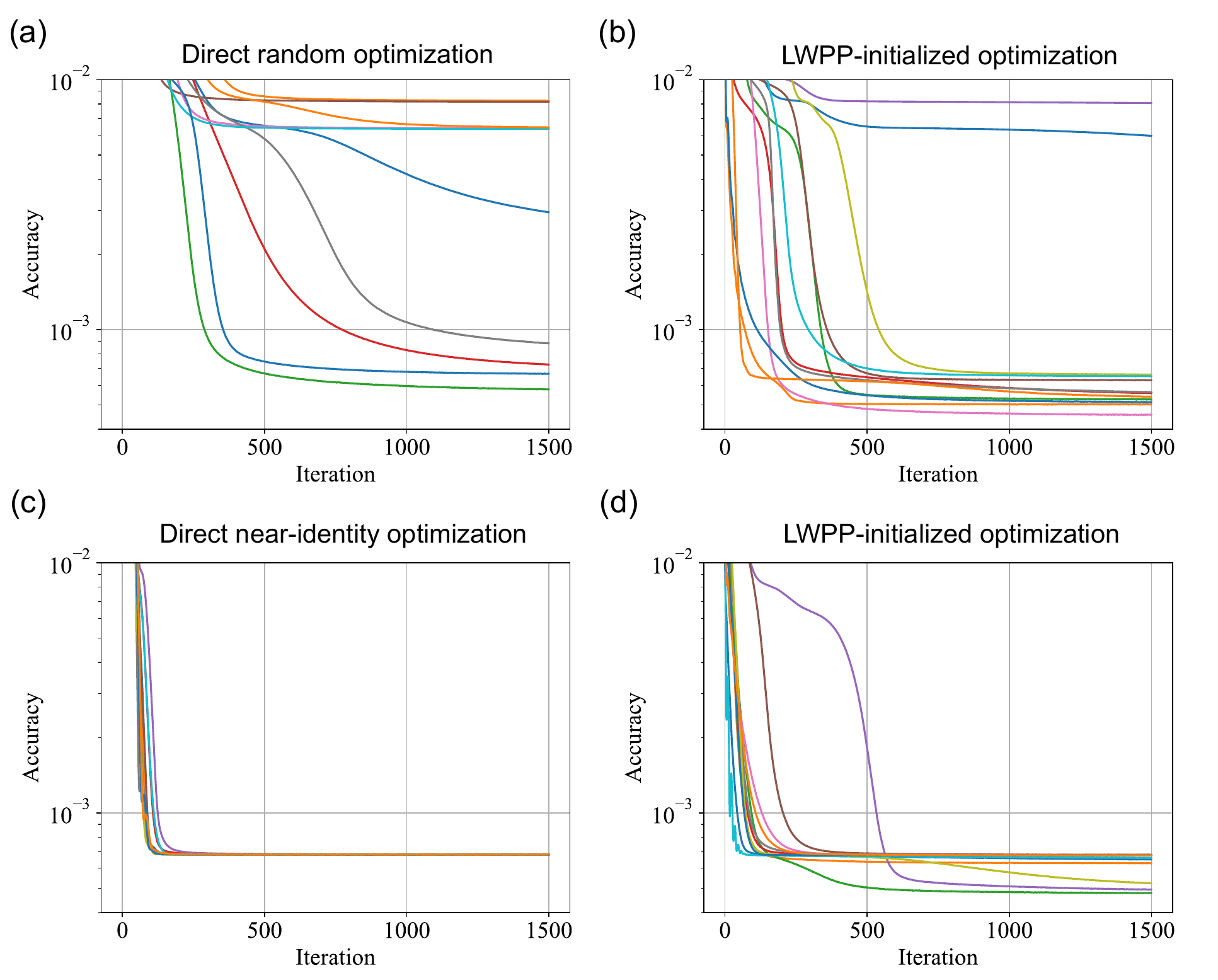}
    \caption{\textbf{LWPP-initialized optimization performance for finding the ground state of a water molecule.} Comparison of VQE optimization trajectories for a circuit with depth $d=4$. Each trace represents an independent optimization run.
    \textbf{(a)} Direct optimization initiated with random parameters, converging to an accuracy of $\approx 6 \times 10^{-4}$.
    \textbf{(b)} LWPP-initialized optimization starting from random parameters, showing faster convergence and achieving a final accuracy below $5 \times 10^{-4}$.
    \textbf{(c)} Direct optimization initiated with near-identity parameters, converging to $\approx 7 \times 10^{-4}$.
    \textbf{(d)} LWPP-initialized optimization starting from near-identity parameters also demonstrates a significant speedup and improved final accuracy compared to the direct method.
    }
    \label{fig:h2o_results}
\end{figure*}

We adapted the VQE ansatz for this problem. The initialization block \(U_i\) was simplified to only use the initial X gates. For the variational block \(U_v\), the two-qubit gates were applied between all adjacent qubits. The circuit depth was set to \(d=4\). The results, comparing direct optimization with LWPP-initialized optimization, are shown in Fig.~\ref{fig:h2o_results}.

Direct optimization starting from random parameters (a) and near-identity parameters (c) converges to accuracies of about \(6 \times 10^{-4}\) and \(7 \times 10^{-4}\), respectively. In this case, direct random initialization performs slightly better than the near-identity heuristic. However, both direct methods are significantly outperformed by the LWPP-initialized runs.

As shown in (b) and (d), LWPP-initialized optimization leads to a clear improvement in both convergence speed and final accuracy. Both runs achieve an accuracy below \(5 \times 10^{-4}\), regardless of whether they started from random or near-identity parameters. This test on a molecular ground state problem confirms that the benefits of LWPP initialization are not specific to the spin model, demonstrating its potential as a broadly useful technique for improving VQA performance.

\subsection{Robustness Analysis: Performance under Quantum Noise}
\begin{figure*}[htbp]
    \centering
    \includegraphics[width=1.0\linewidth]{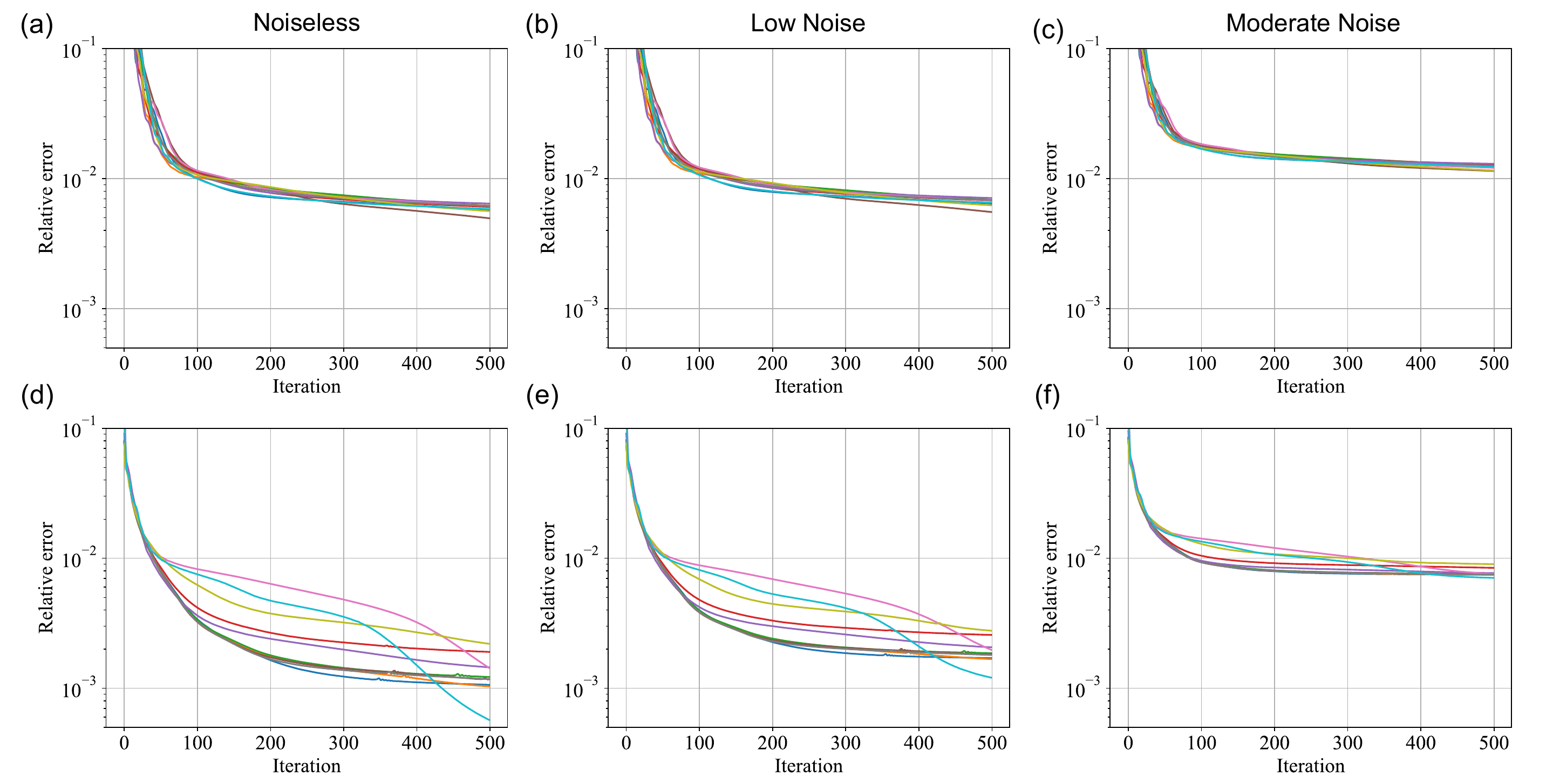} 
    \caption{\textbf{Robustness of LWPP initialization under quantum noise.} 
    Comparison of optimization trajectories for the antiferromagnetic Heisenberg XYZ model ($3 \times 3$ lattice, $d=6$) under a depolarizing noise model with probability $p$.
    \textbf{Top Row (a-c): Direct optimization} starting from near-identity parameters. 
    \textbf{Bottom Row (d-f): LWPP initialization} seeded by pre-optimization.
    (a, d) \textbf{Noiseless ($p=0$):} LWPP demonstrates a massive advantage.
    (b, e) \textbf{Low Noise ($p=10^{-6}$):} The advantage of LWPP is preserved, with trajectories converging significantly deeper than direct optimization.
    (c, f) \textbf{Moderate Noise ($p=10^{-5}$):} While noise degrades absolute performance, LWPP (f) consistently reaches sub-$10^{-2}$ accuracy, whereas direct optimization (c) stagnates above $10^{-2}$. Notably, LWPP matches the final 500-step direct result in only dozens of iterations, demonstrating that its landscape-navigation advantage provides both better precision and massive speedup under realistic noise. In all noisy scenarios, both the gradients used during optimization and the final relative errors are evaluated in the presence of noise.}
    \label{fig:noise_robustness}
\end{figure*}
In practical noisy intermediate-scale quantum applications, hardware noise is unavoidable. To investigate the robustness of our initialization strategy, we benchmarked the performance on the antiferromagnetic Heisenberg XYZ model on a $3 \times 3$ lattice with a circuit depth of $d=6$. We adopted a standard local depolarizing noise model, where a symmetric depolarizing channel is applied to the relevant qubits immediately after every single-qubit and two-qubit gate.

The channel is parameterized by the single probability parameter $p$ (where $p_x=p_y=p_z=p$), resulting in a total error probability per gate of $P_{err} = 3p$. We tested three regimes: noiseless ($p=0$), low or moderate noise ($p=10^{-6}, 10^{-5}$). During the optimization, gradients and energy costs are evaluated under the influence of noise. The comparative results are presented in Fig.~\ref{fig:noise_robustness}.

\begin{itemize}
    \item \textbf{Noiseless Baseline (Fig.~\ref{fig:noise_robustness}a vs d):} In the ideal case ($p=0$), LWPP-initialized optimization (d) clearly avoids the local minima that trap the direct optimization (a), achieving a significantly lower relative error (reaching $\sim 10^{-3}$).
    
    \item \textbf{Low-Noise Regime (Fig.~\ref{fig:noise_robustness}b vs e):} With a noise parameter $p=10^{-6}$, the advantage of LWPP is fully preserved. While direct optimization (b) stagnates around $10^{-2}$, the LWPP-initialized runs (e) consistently converge to much deeper minima.
    
    \item \textbf{Moderate-Noise Regime (Fig.~\ref{fig:noise_robustness}c vs f):} Even when the noise increases, the advantage of LWPP initialization remains significant. In this regime, the performance of direct optimization (c) degrades considerably, with relative errors stagnating above $10^{-2}$ throughout the 500 iterations. In contrast, LWPP-initialized runs (f) consistently drive the relative error below the $10^{-2}$ threshold. Remarkably, LWPP matches the 500-step direct results in only dozens of iterations.
\end{itemize}

These results challenge the notion that noise might render advanced initialization unnecessary by washing out the landscape. Instead, we observe that the two methods are constrained by different factors. Direct optimization remains primarily limited by \textit{landscape traps}, often getting stuck in high-energy local minima. In contrast, LWPP initialization, by smoothing the search space, helps the optimizer bypass these traps even under decoherence. This suggests that while noise reduces overall fidelity, it does not eliminate the optimization ruggedness that hinders direct methods. Consequently, the smoothed basin provided by $E_{\text{LWPP}}$ remains a necessary guide for reaching better solutions in realistic VQA scenarios.

In summary, LWPP initialization is not merely a tool for ideal simulations; it is also a reliable strategy for noisy hardware. While noise inevitably degrades the absolute performance of both initialization strategies, LWPP-initialized optimization consistently maintains a significant advantage margin over the direct approach, ensuring superior accuracy and efficiency in realistic environments.

\subsection{Parameter Correlations Beyond Probability Distributions}
A crucial question regarding the effectiveness of LWPP initialization is what precisely the classical pre-optimization phase is learning. Is it merely identifying a more favorable statistical distribution of parameters from which any random sample would be a good starting point, or is it finding a specific, structured set of parameters whose inter-correlations are critical? This section provides strong evidence for the latter.

\begin{figure*}[t!]
    \centering
    \includegraphics[width=0.8\textwidth]{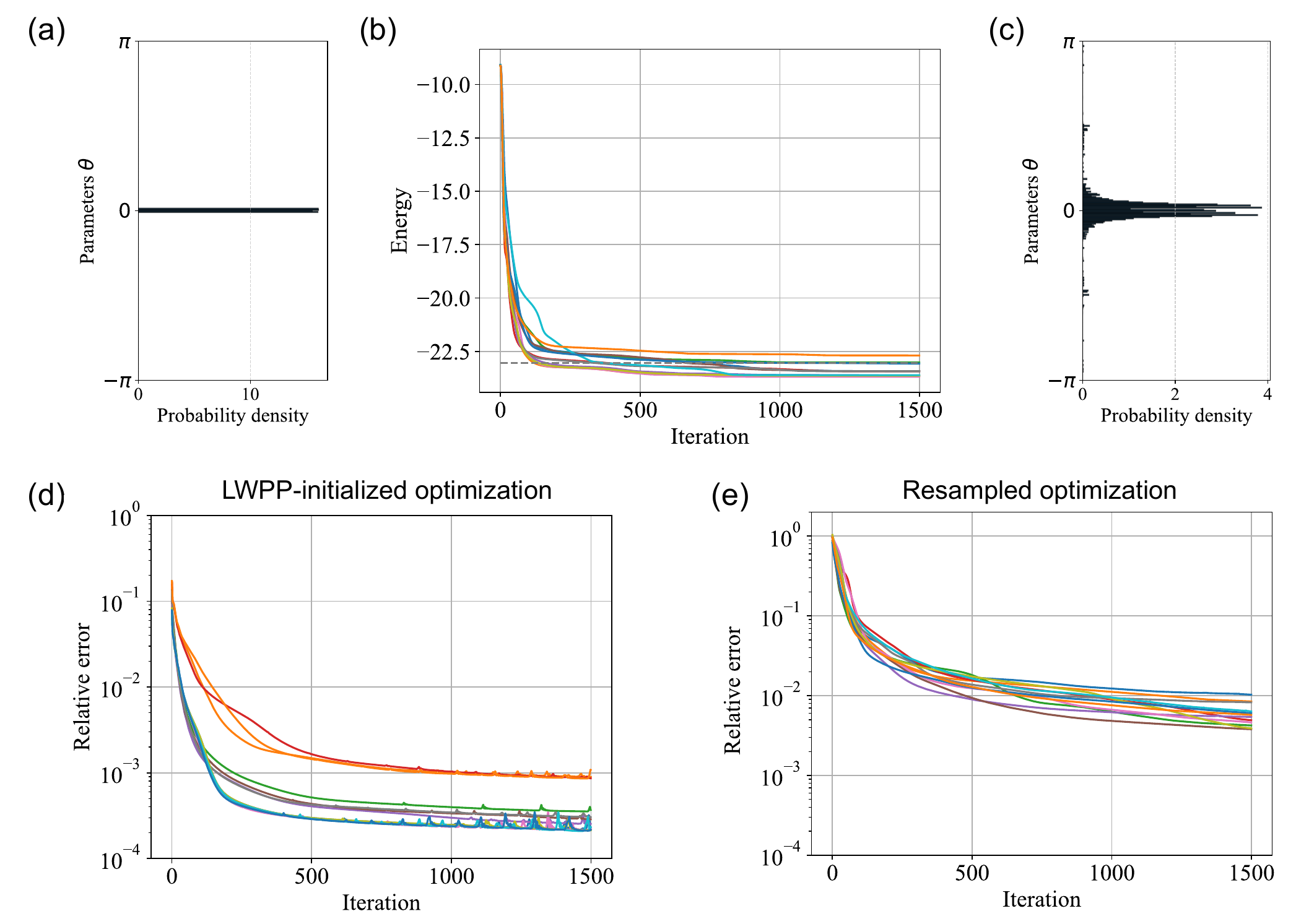}
    \caption{\textbf{LWPP learns parameter correlations, not just a statistical distribution.} 
    Results for a VQE optimization on a $3 \times 4$ lattice with circuit depth $d=6$. 
    (a) The initial near-identity distribution of circuit parameters $\theta$, sharply peaked at zero. 
    (b) LWPP pre-optimization trajectory starting from the near zero parameters, with the initial probability distribution shown in (a).
    (c) The parameter distribution after it has been evolved by the classical LWPP pre-optimization. The distribution is now broader.
    (d) The VQA optimization trajectory with LWPP initialization (initialized with parameters drawn from the end of (b)). The optimization is highly successful, achieving a final accuracy near $10^{-4}$.
    (e) The VQA optimization accuracy when initialized with a new set of parameters resampled from the distribution shown in (c). The loss of the specific parameter correlations results in a significantly degraded final accuracy, which is more than an order of magnitude worse.
    }
    \label{fig:parameter_correlations}
\end{figure*}

We demonstrate this with a representative example on an antiferromagnetic Heisenberg XYZ model with $3 \times 4$ lattice using a VQA ansatz with circuit depth $d=6$, as illustrated in Fig.~\ref{fig:parameter_correlations}. The process begins by initializing the circuit with parameters drawn from a near-identity distribution, where all parameters $\theta$ are sharply peaked around zero, as shown in Fig.~\ref{fig:parameter_correlations}(a). We then perform the classical pre-optimization using the LWPP cost function. The resulting distribution of the optimized parameters is shown in Fig.~\ref{fig:parameter_correlations}(c). As expected, the distribution has broadened, indicating that the optimizer has moved the parameters away from the strict near-identity configuration to minimize the LWPP cost.

The critical test lies in how these parameters are used. First, we take the exact parameter set, $\boldsymbol{\theta}_{\text{LWPP}}$, found by the pre-optimizer and use it to initialize a standard VQA optimization. The results, shown in Fig.~\ref{fig:parameter_correlations}(b) and (d), are excellent. The optimization converges rapidly to a high-fidelity solution, achieving a final accuracy approaching $10^{-4}$.

Second, we conduct a control experiment. We treat the final parameter distribution from Fig.~\ref{fig:parameter_correlations}(c) as a statistical model and generate a new set of parameters, $\boldsymbol{\theta}_{\text{resampled}}$, by resampling from it. While this new set has the same overall statistical properties, it has lost the specific correlations and structure present in $\boldsymbol{\theta}_{\text{LWPP}}$. When this resampled set is used to initialize the VQA, the performance is dramatically degraded, as shown in Fig.~\ref{fig:parameter_correlations}(e). The final accuracy is more than an order of magnitude worse than in the direct-use case.

This significant performance gap unequivocally demonstrates that LWPP initialization does not simply learn a better probability distribution. Instead, it uncovers the intrinsic, high-dimensional correlations between parameters that define a favorable starting point. The specific structure of the parameter set is the key to its success, acting as a navigator that places the optimizer in a promising basin of attraction that is otherwise difficult to find.

\subsection{Efficiency under High Landscape Ruggedness}
\begin{figure*}[t!]
    \centering
    \includegraphics[width=0.8\textwidth]{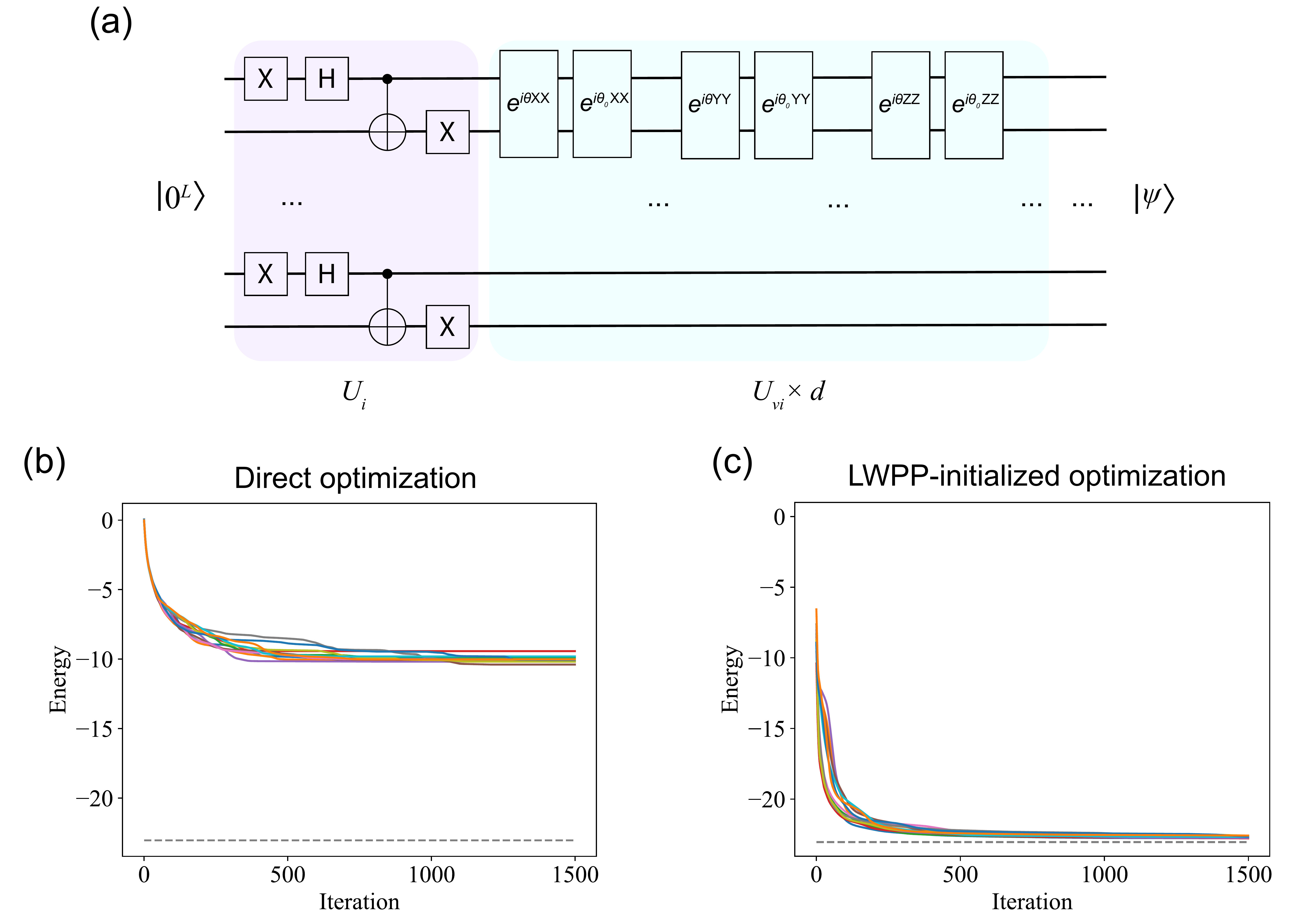}
    \caption{\textbf{LWPP initialization demonstrates robustness in a rugged landscape.}
    \textbf{(a)} The modified VQE ansatz, where an additional fixed-angle, random-parameter rotation gate ($e^{i\phi P}$) is inserted after each trainable gate ($e^{i\theta P}$). This makes the landscape near the identity highly non-trivial.
    \textbf{(b)} Direct VQA optimization, initiated with near-identity parameters ($\theta \approx 0$), consistently fails to find the ground state (dashed line), becoming trapped in high-energy local minima due to the rugged landscape.
    \textbf{(c)} In stark contrast, the LWPP-initialized runs, which also begin their pre-optimization from near-identity parameters, successfully converge to the ground state energy. This highlights the power of the LWPP landscape as a robust navigator.
    }
    \label{fig:rugged_landscape}
\end{figure*}

To further probe the advantages of LWPP-initialized optimization, we designed a more challenging benchmark that intentionally creates a rugged and non-trivial optimization landscape, even near the identity. In this scenario, we modify the VQE ansatz by inserting an additional fixed-parameter rotation gate after each trainable gate, as shown schematically in Fig.~\ref{fig:rugged_landscape}(a). Specifically, after each parameterized gate $e^{-i\theta P}$, we apply a gate $e^{-i\theta_0 P}$ of the same type, where the angle $\theta_0$ is chosen from a fixed, random distribution. This modification disrupts the smooth path from the identity circuit to the solution, providing a stringent test for any initialization heuristic.

We compare two strategies. The first is a standard VQA optimization initialized with near-identity parameters ($\theta \approx 0$). As shown in Fig.~\ref{fig:rugged_landscape}(b), this direct approach completely fails. The presence of the fixed random rotations creates a complex local energy landscape that immediately traps the optimizer. Despite multiple independent runs, none are able to find a path to a low-energy solution, instead converging to various high-energy local minima.

The second strategy is our proposed LWPP-initialized optimization. Crucially, we still begin the classical pre-optimization phase with near-identity parameters ($\theta \approx 0$). The LWPP cost function therefore must contend with the full, complex circuit structure. The results, shown in Fig.~\ref{fig:rugged_landscape}(c), are positive. The LWPP-initialized runs consistently and rapidly converge to the true ground state energy (dashed line).

This outcome demonstrates the core thesis of our work. The LWPP cost function, while numerically approximate, provides a superior and robust ``navigator" for the optimization landscape. It is able to effectively find a structured set of initial parameters $\boldsymbol{\theta}$ that counteracts the effect of the fixed random rotations, placing the VQA in a promising basin of attraction. This test confirms that LWPP initialization is not merely a heuristic for ideal circuits but a powerful tool capable of overcoming significant landscape ruggedness where standard methods break down.

\section{Optimization Workflows and Simulation Details}
\label{sec:details}

\subsection{Summary of Optimization Workflows}
\begin{figure}[h]
    \centering
    \includegraphics[width=0.8\linewidth]{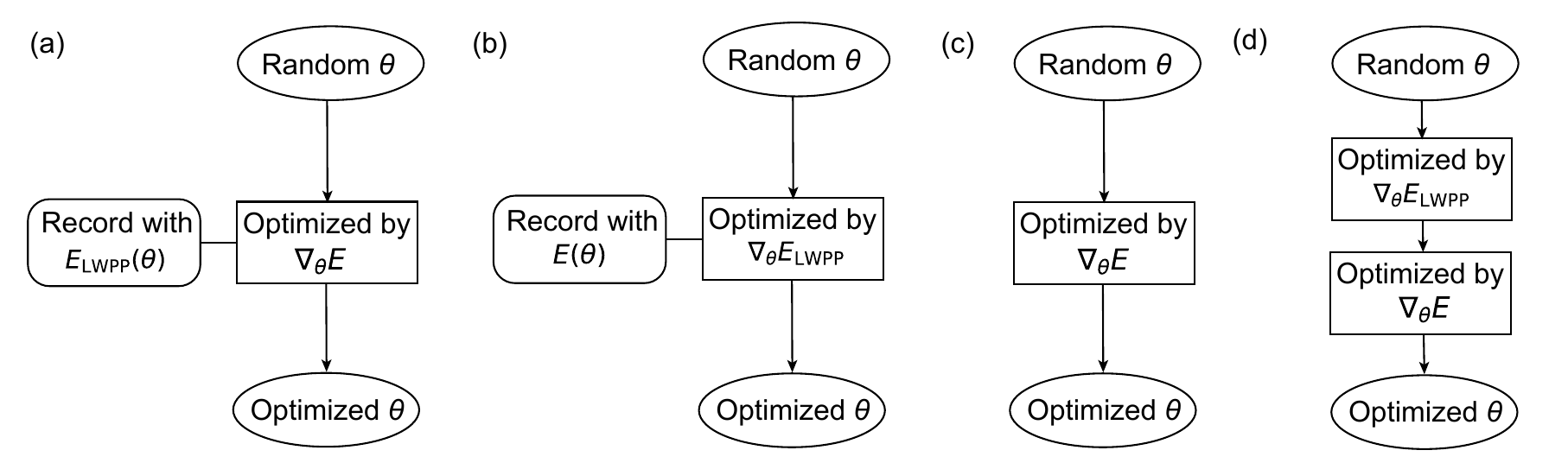} 
    \caption{\textbf{Schematic overview of the optimization workflows.} 
    (a) Benchmarking LWPP as a simulator by evaluating its estimation accuracy along a directed exact optimization trajectory. 
    (b) Validating LWPP as a navigator by assessing whether optimization on the approximate landscape effectively guides parameters toward high-quality regions. 
    (c) Standard direct VQA optimization starting from heuristic initialization. 
    (d) Our proposed LWPP-initialized optimization strategy, featuring a classical pre-optimization phase to navigate toward the global basin of attraction before performing exact quantum fine-tuning.}
    \label{fig:workflow_summary}
\end{figure}

To clarify the methodological details and distinct usage scenarios discussed in the main text, we summarize the four algorithmic workflows in Fig.~\ref{fig:workflow_summary}.

\begin{itemize}
    \item \textbf{(a) Exact optimization + LWPP evaluation:} This protocol is used to benchmark the estimation accuracy. The optimizer updates parameters using the exact energy gradients $\partial_\theta E$, while $E_{\text{LWPP}}$ is recorded passively to monitor its deviation from the true energy.
    
    \item \textbf{(b) LWPP optimization + Exact evaluation:} This protocol validates the navigation capability of the LWPP landscape. The optimizer is driven by the approximate gradients $\partial_\theta E_{\text{LWPP}}$. We record the exact energy $E(\theta)$ to verify if minimizing the LWPP cost effectively guides parameters to low-energy regions.
    
    \item \textbf{(c) Direct optimization:} The standard baseline strategy where variational parameters are optimized directly on the exact energy landscape starting from random initialization.
    
    \item \textbf{(d) LWPP-initialized optimization (proposed strategy):} Our proposed two-stage method. In the first phase, parameters are pre-optimized using $\partial_\theta E_{\text{LWPP}}$ to reach a favorable basin of attraction. In the second phase, these parameters serve as the seed for the final optimization using $\partial_\theta E$ to achieve high precision.
\end{itemize}

\subsection{VQE Ansatz}
\begin{figure}[h]
\centering
\includegraphics[width=0.7\linewidth, keepaspectratio]{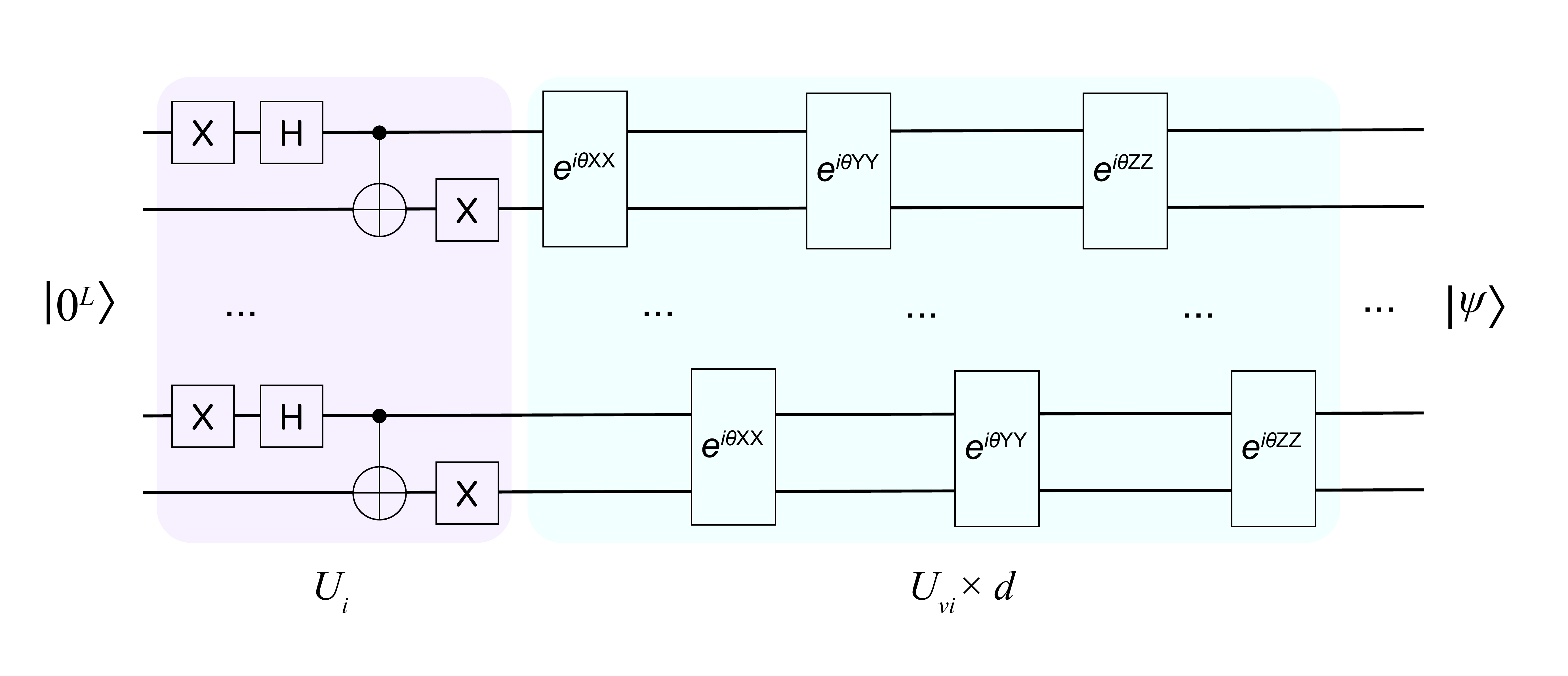}
\caption{\textbf{VQE ansatz employed in the main text.}
    The circuit begins with an initialization block, $U_i$, that prepares an entangled state of Bell pairs from the initial state $|0^L\rangle$. Subsequently, the variational block, $U_v$, applies layers of parameterized nearest-neighbor rotations ($e^{i\theta XX}$, $e^{i\theta YY}$, $e^{i\theta ZZ}$) corresponding to the 2D lattice of the Heisenberg XYZ model. This block is repeated $d$ times with independent parameters to prepare the final trial state $|\psi\rangle$. Ellipses (...) indicate omitted qubits and gates for clarity.
    }
\label{fig:ansatz}
\end{figure}

The VQE trial states used in this work are prepared by the hardware-efficient parameterized quantum circuit (PQC) shown schematically in Fig.~\ref{fig:ansatz}, such that $|\psi(\boldsymbol{\theta})\rangle = U(\boldsymbol{\theta})|0^L\rangle$. The circuit begins with a fixed initialization block, $U_i$, which applies a layer of gates (including Hadamard, X, and CNOTs) to create an entangled state of Bell pairs, $1 / \sqrt{2}(|01\rangle-|10\rangle)$, on neighboring qubits so that the initial state is in the $J_{\mathrm{tot}} = 0$ sector. This is followed by a parameterized variational block, $U_v$, which is repeated $d$ times, where $d$ is the circuit depth. Each variational block $U_v(\boldsymbol{\theta}_i)$ is designed to match the structure of the 2D Heisenberg XYZ model and consists of layers of nearest-neighbor two-qubit rotation gates, given by:
\begin{equation}
\label{eq:variational_block}
    U_v(\boldsymbol{\theta}_i) = 
    \left( \prod_{\langle j,k \rangle} e^{-i \theta^{(i)}_{jk,Z} Z_j Z_k} \right)
    \left( \prod_{\langle j,k \rangle} e^{-i \theta^{(i)}_{jk,Y} Y_j Y_k} \right)
    \left( \prod_{\langle j,k \rangle} e^{-i \theta^{(i)}_{jk,X} X_j X_k} \right).
\end{equation}
Here, the notation $\langle j,k \rangle$ denotes a pair of nearest-neighbor qubits on the 2D square lattice. For each gate type (XX, YY, or ZZ), the operators corresponding to different bonds commute with each other. Therefore, the order of application for the gates within each product $\prod_{\langle j,k \rangle}$ is arbitrary. The full PQC is constructed by composing these blocks, $U(\boldsymbol{\theta}) = (\prod_{i=1}^d U_v(\boldsymbol{\theta}_i)) U_i$, where each of the $d$ blocks has its own independent set of tunable parameters $\boldsymbol{\theta}_i$. For a lattice of size $m \times n$, each of the three gate layers contains $2mn$ rotation gates. All numerical simulations presented are ideal, omitting hardware noise and shot-sampling error in the main text. Gradients are computed via automatic differentiation using the TensorCircuit-NG framework~\cite{zhang2023tensorcircuit}. The parameters are optimized using the Adam optimizer with a learning rate of 0.01.

\end{document}